\documentclass[preprint,review,12pt]{elsarticle}

\usepackage{amssymb}
\usepackage{subfigure}
\usepackage{graphicx}
\usepackage[T1]{fontenc}
\usepackage{python}

\journal{Journal of Molecular Liquids}

\begin{document}

\begin{frontmatter}



\title{Stripes polymorphism and water-like anomaly in hard core-soft corona dumbbells}


\author[inst1]{T. P. O. Nogueira}

\affiliation[inst1]{organization={Departamento de F\'{i}sica, Instituto de F\'{i}sica e Matem\'{a}tica, 
Universidade Federal de
Pelotas},
            addressline={Caixa Postal 354}, 
            city={Pelotas},
            postcode={96001-970}, 
            state={Rio Grande do Sul},
            country={Brazil}}

\author[inst1]{José Rafael Bordin}


\begin{abstract}
In this paper we investigate the phase diagram of a dumbbell model composed of two hard-core soft-corona beads through $NpT$ simulations. This particular system was chosen due to its ability to exhibit a diverse range of stripe patterns. Analyzing the thermodynamic and structural changes along compression isotherms, we explore the transition between these distinct patterns. In addition to the stripe and Low-Density-Triangular solid phases obtained, we observed a Nematic Anisotropic phase characterized by a polymer-like pattern at high temperatures and intermediate pressures. Furthermore, we demonstrate the significant role played by the new characteristic length scale, which arises from the anisotropic geometry of the dumbbell structure, in the transition between the stripes patterns. Notably, not only do the structural properties exhibit intriguing behavior, but the diffusion and density in the nematic fluid phase also displays a water-like anomalous increase under compression. Those findings can be valuable in guiding the design of materials based on nanoparticles, with the aim of achieving specific mesopatterns.
\end{abstract}




\end{frontmatter}


\section{Introduction}

Lately the scientific community is devoting theoretical~\cite{Mambretti21,Cardoso2021,nogueira2022patterns,baran2022influence,lindsay2019equilibrium,krook2019experiments} and experimental~\cite{li2022geometry, zenati2022triblock,pula2022solvent,abate2016shear,amadi2022nanoscale} studies to understand the physics and develop new materials by the means of the Self-assembly process. That refers to the spontaneous aggregation and organization of molecules, resulting in a wide range of structures. This natural process is observed in various biological processes and serves as an effective approach for creating ordered nanostructures. The goal is to produce new materials with exceptional properties~\cite{Vogel15,Brinker1999,amadi2022nanoscale,pula2022solvent}. Building blocks as block of copolymers (BCP) or other colloidal particles whose interaction has a long-range repulsion  with a very short-ranged attractive scale can assembly into polymer-like mesophases and distinct stripes patterns~\cite{Haddadi2021, Haddadi2021a} and has shown applications in optical coatings~\cite{vignolini2012k,hulkkonen2018all}, metamaterials~\cite{kilchoer2020strong,murataj2021hyperbolic,alvarez2021block}, photonic crystals~\cite{noro2016enthalpy,liberman2017application}, plasmonic nanostructures~\cite{mistark2009block,rahman2014block} and membranes~\cite{abetz2015isoporous,nunes2016block}.

Rigorously, an optimum assembly enables the system to become thermodynamic stable, giving rise to well-defined structures. However, due to unexpected intermolecular interactions, initial state conditions, kinetic effects and so forth the assembly may fall into a metastable state~\cite{chen2023catalytic}. When treating BCP this feature is frequently encountered. Therefore, structures such as stripes patterns are observed. These patterns are formed but not only by highly confined, single-layer-thick films of cylinder-forming~\cite{hahm1998defect,hahm2000cylinder,hahm2001time} or by compositionally symmetric, lamella-forming~\cite{ouk2003epitaxial,campbell2012network,campbell2013processing,campbell2013topologically,kim2009spontaneous} BCPs. These stripes patterns are not free of defects and it was pointed out that those are not equilibrium fluctuations but long-lived metastable states into which the kinetics of structure formation have been trapped~\cite{li2015defects}. 

The stripes pattern was also observed in computer simulations ranging from quantum to classical approaches~\cite{nogueira2022patterns,chen2023catalytic,clark2009hexatic,hertkorn2021pattern,Pattabhiraman2017}, including recent studies involving competitive systems~\cite{bordin2019distinct,mendoza2019mechanism,mendoza2021cluster,mendoza2022exploring,malescio2022self, Cardoso2021, Bordin2018,Bordin2018a}. Colloidal suspensions, as BCP in a solvent, are a good example of such systems. Due to the characteristics of BCP, each colloid will be made of molecular subunits which form a central packed agglomeration and a less dense and more entropic peripheral area. These distinct conformations compete to rule the suspension behavior. Therefore, we can model the competitions in a system by modeling the colloids interaction using a Hard-Core Soft-Corona (HCSC) potential with two length scales. Such approach has been employed since the so-called ramp-like Jagla potential~\cite{Ja98, Jagla1999a, jagla1999b}. The HCSC interaction potentials can exhibit many shapes~\cite{Buldyrev2009}: square shoulder, linear, soft or convex ramp, quasi-exponential tail and inverse power potentials~\cite{Malescio2005,Barbosa2013,franzese2010,DeOliveira2006,BarrosdeOliveira2007,Malescio2011,Yan2005,Contreras-Aburto2010,Coslovich2013,Saija2009, Nowack2019, Pineros16,Malescio2021,Prestipino2010, Das13,Malescio11}. What they share in common is the competition in between two particles conformations: (i) a soft repulsion at long range distance $r_2$; and (ii) a hard-core repulsion at shorter distances $r_1$~\cite{Marques21b, Marques2020}. Even for purely repulsive potentials, the formation of stripes is guided by the desire to minimize the potential energy of the system~\cite{Pattabhiraman2017}. 

In a purely repulsive approach~\cite{Bordin2018, Cardoso2021}, each colloid has overlapping coronas with two neighboring particles along the stripe. However, the stripes are aligned in such a way that the distance between them exceeds the size of the corona. As a result, there is no overlap between the coronas of adjacent stripes. This arrangement is more energetically favorable compared to a structure where particles are equally spaced, and each particle's corona partially or fully overlaps with all of its neighboring particles. Most recently~\cite{nogueira2022patterns}, we have shown that by introducing anisotropy to a HCSC competitive system, distinct stripes patterns can be obtained. As a consequence of an extra length scale introduced by the dumbbell shape of the molecules. This creates and extra competition in the system, leading to unique assembly patterns. Furthermore, it has been widely recognized that the interplay between the two length scales in core-softened potentials can give rise to the presence of anomalies reminiscent of those observed in water~\cite{yan06,gibson2006,franzese2007,Xu11,bordin15, bordin16,Krott2016, haro18,saki18, Ryzhov20, roca22,bretonnet22}. 

Now, we revisit the dumbbells case~\cite{nogueira2022patterns} to extend our study and explore the low temperature phase diagram. Our goal is to check the stability of the anisotropy-induced patterns upon heating and compressing. To this end, extensive $NpT$ simulations were carried out for dimers with fixed intramolecular separation. Our results not only unveil details about the structural transitions between the distinct stripes patterns, but also show the existence of a reentrant nematic anisotropic fluid phase with polymer-like structure at high temperatures and intermediate pressures. The reentrant phase pressure range coincides with the anisotropy-induced patterns region. Finally, this fluid phase exhibits water-like diffusion, characterized by a maxima in the self-diffusion coefficient under compression along a isotherm, and density anomaly, defined by a temperature of maximum density along a isobar.

The paper is organized as follow. In Section~\ref{sec:model} we show the HCSC model employed, the simulation details and the quantities measured to analyze the system. Followed by the results and their discussion in Section~\ref{sec:results} and finally our conclusion in Section~\ref{sec:conclusion}.

\section{Model and Simulation Details}\label{sec:model}
\subsection{The Model}

 To explore the low temperature behavior we carried out 2D $NpT$ simulations with periodic boundary conditions using the Sandia National Laboratories' LAMMPS\cite{Plimpton1995}. The intermolecular HCSC interaction was modeled by the core-softened potential composed by a short-range attractive Lennard-Jones potential plus a repulsive Gaussian term centered at $r_0$, with depth $u_0$ and width $c_0$.
\begin{equation}
        U_{CS}(r) = 4\epsilon \bigg[\bigg(\frac{\sigma}{r}\bigg)^{12} - \bigg(\frac{\sigma}{r}\bigg)^{6}  \bigg] +
        u_0 \exp\bigg[-\frac{1}{c_{0}^{2}}\bigg(\frac{r-r_0}{\sigma}\bigg)^2\bigg].
        \label{eq:CS}
\end{equation}

\begin{figure}[ht]
    \centering
    \includegraphics[width=0.75\textwidth]{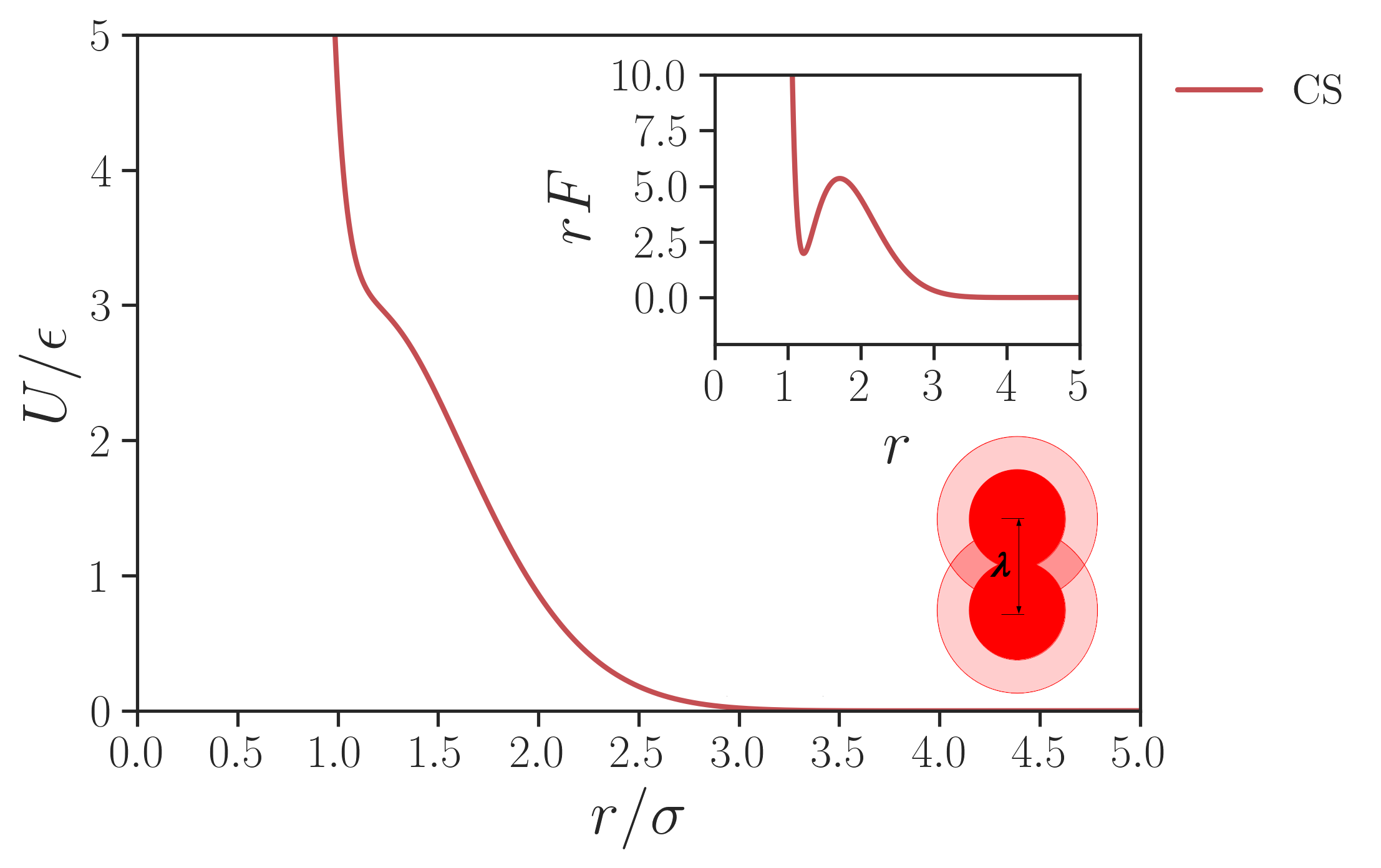}
    \caption{Core-softened potential. Graph inset: the Core-Softened force times the length scale. Dumbbell inset is the illustration for the dimeric molecules with intramolecular separation given by the $\lambda$ parameter.}
    \label{fig:potential}
\end{figure}

Using the parameters $u_0 = 5\epsilon$, $c_0^2 = 1.0$, and $r_0/\sigma = 0.7$~\cite{Bordin2018,BarrosDeOliveira2006,DeOliveira2006} the potential \ref{eq:CS} exhibits a ramp-like shape as shown in Fig.\ref{fig:potential}. As de Oliveira and co-authors have shown~\cite{BarrosDeOliveira2010}, both the real and the imaginary branch of the instantaneous normal mode (INM) spectra of this potential have a pronounced bimodality which must be connected with two different length scales -- unlike simple liquids, as Lennard Jones or Morse fluids, who have only one scale. The first length-scale, correspondent to the hard-core, is located near $r_1 = 1.2\sigma$, where the force has a local minimum~\cite{BarrosDeOliveira2010} (see Fig.~\ref{fig:potential} graph inset), while the longer length scale, the soft corona, is located at $r = 2.0\sigma$, where the fraction of imaginary modes of the INM spectra has a local minimum and a maximum is expected in the radial distribution function~\cite{BarrosDeOliveira2010}. The cutoff radius for the interaction is $r_c = 3.5\sigma$.

\subsection{Simulation Details}

Our simulations were carried out in the isothermal-isobaric ensemble ($NpT$) with fixed number $N_{tot} = 2000$ of dumbbell colloids. The separation between two monomers in the same dumbbell is $\lambda = 0.50$ and was kept rigid by the SHAKE algorithm~\cite{Ryckaert1977}. This value was chosen based in our previous work~\cite{nogueira2022patterns} once it shows distinct stripes phases. The temperature and pressure were controlled using the Nose-Hoover thermostat and barostat. The thermostat damping parameter is $Q_T = 0.1$, and for the barostat $Q_p = 1000$, respectivelly. The colloids were initially randomly placed in a dilute state with molecule number density $\rho_0 = N_{tot}/A_0 = 0.01$, where $A_0 = L_0\times L_0$ is the initial box. Then, $8 \times 10^6$ time steps were performed to equilibrate the system, followed by $1 \times 10^7$ time steps for the results production stage, with averages and snapshots being taken at every $1 \times 10^4$ steps. The time step is $\delta t = 0.001$. To ensure that the system temperature and pressure were well controlled we averaged the instantaneous value of those quantities during the simulations. As well, to ensure that the system reached the equilibrium, we evaluated the kinetic and the potential energy along the simulation. Since they did not vary with time, only oscillated around a mean value after a long simulation time, we assumed that the system was in equilibrium~\cite{Bordin2018a}. The colloid number density $\rho$ along a simulations is defined as $\rho = N_{tot} / <A_m>$, with $<A_m>$ being the mean area at a given temperature and pressure. Isotherms were evaluated in the interval of $T = 0.01$ to $T = 0.2$ with temperature increase of $\delta T = 0.01$. The pressure was varied from $p = 0.05$ up to $p = 15.00$. The isothermal compressibility, $K_T= \frac{1}{\rho} \left ( \frac{\partial \rho}{\partial P} \right )_T$, was then evaluated under a compression path. Once we did not evaluate $K_T$ under a decompression path, we stress that the point of transition can suffer a shift due to hysteresis.

The colloids translational structure was analyzed by the radial distribution function (RDF) $g(r)$. Once the interactions are pairwise, important quantities can be calculated explicitly as integrals involving the RDF~\cite{hansen1990theory}. Nonetheless, the two-body contribution $s_2$ to the entropy can be directly calculated from~\cite{Baranyai1989}
\begin{equation}
s_2 = -\frac{\rho}{2}\int[g(r_{ij})\ln(g(r_{ij})) - g(r_{ij}) +1]d{\bf r}\;.
\label{eq: s2}
\end{equation}
The two-body excess entropy is a structural order metric that allow us to correlate thermodynamics and structure, but it is not a thermodynamic property of the system. However, it is a powerful tool to analyze structural characteristics of the core-softened system, as the water-like structural anomaly~\cite{Sharma2006, Lascaris2010,Krott2015}: in normal fluids, it decrease under compression, while an increase indicates an anomalous behavior. To depict the long range translational ordering using the RDF as basis we evaluate the cumulative two-body entropy~\cite{Klumov20, Cardoso2021}, 
\begin{equation}
    C_{s_2} (R) = -\pi \int_{0}^{R} [g(r_{ij})\ln(g(r_{ij})) - g(r_{ij}) +1] r_{ij} d r_{ij}.
\end{equation}
Here $r_{ij}$ is the distance and $R$ is the upper integration limit. At this distance $|C_{s_2}|$ converges for disordered phases and diverges for ordered phases. 

The orientational order is checked using the bond orientational order parameter $\Psi_l$,

\begin{equation}
\label{ppsi6}
\Psi_l = \frac{1}{N} \sum_{m=1}^N \psi_l(r_m)
\end{equation}
\noindent where

\begin{equation}
\label{psi6}
\psi_l(r_{m}) = \frac{1}{n_N} \sum_{n=1}^{n_N} \exp[li\theta_{mn}]\;.
\end{equation}
is the local bond orientational order parameter. The sum $n$ is over all the $n_N$ nearest neighbors of $m$ - the neighboring particles were picked by Voronoi tesselation~\cite{Prestipino2012}. $\theta_{mn}$ is the angle between some fixed axis and the bond joining the $m-th$ particle to the $n-th$ neighboring particle. For a triangular lattice, $l = 6$ and $|\Psi_6| \rightarrow 1.0 $  if the colloids are in a perfect triangular lattice. On the other hand, for the stripe phase, a typical structure in simulations of core-softened fluids and experiments for colloidal films~\cite{Fomin2019}, we consider $l = 2$ to analyze the twofold stripe order, as proposed by Hurley and Singer~\cite{Hurley1992}. The orientational correlation given by 
\begin{equation}
    \label{eq: gs-corr}
    g_l (\mathbf{r}) = \langle \psi_l (\mathbf{r}) \psi_l^* (\mathbf{0})\rangle, 
\end{equation}
was analyzed to assess the long range orientational ordering. This analysis was carried out using the Freud library~\cite{freud2020}. Also, since we are analyzing the compression along a isotherm, another useful quantity to understand the transition between distinct stripes patterns under compression is the isothermal bond orientational order parameter susceptibility,
\begin{equation}
    \chi_{\Psi_2} = \frac{1}{{k_B}}\left(\frac{\partial \Psi_2}{\partial P}\right)_T \,.
\end{equation}
Here, we $l=2$ was chosen once we are focusing in transitions between phases with twofold order.

The dynamic behavior was analyzed by the colloid center of mass mean square displacement (MSD) given by 
\begin{equation}
    \label{eq: MSD}
    \langle [ \mathbf{r}(t) - \mathbf{r}(t_0) ]^2 \rangle = \langle \Delta \mathbf{r}(t) ^2 \rangle, 
\end{equation}
where $\mathbf{r}(t_0)$ and $\mathbf{r}(t)$ denote the particle position at a time $t_0$ and at a later time $t$, respectively. The MSD is then related to the diffusion coefficient $D$ by the Einstein relation, 
\begin{equation}
    \label{eq: diff}
    D = \lim_{t \rightarrow \infty} \frac{\langle \Delta \mathbf{r}(t) ^2 \rangle}{4t}.
\end{equation}

\section{Results and Discussion}\label{sec:results}

In our recent work~\cite{nogueira2022patterns} we have observed the existence of distinct stripes patterns in colloids dumbbells with $\lambda = 0.50$ along the isotherm $T = 0.10$. Now, we explore the low temperature phase diagram to check the stability of such patterns upon heating.

\begin{figure*}[ht]
    \centering
    \includegraphics[width=0.85\textwidth]{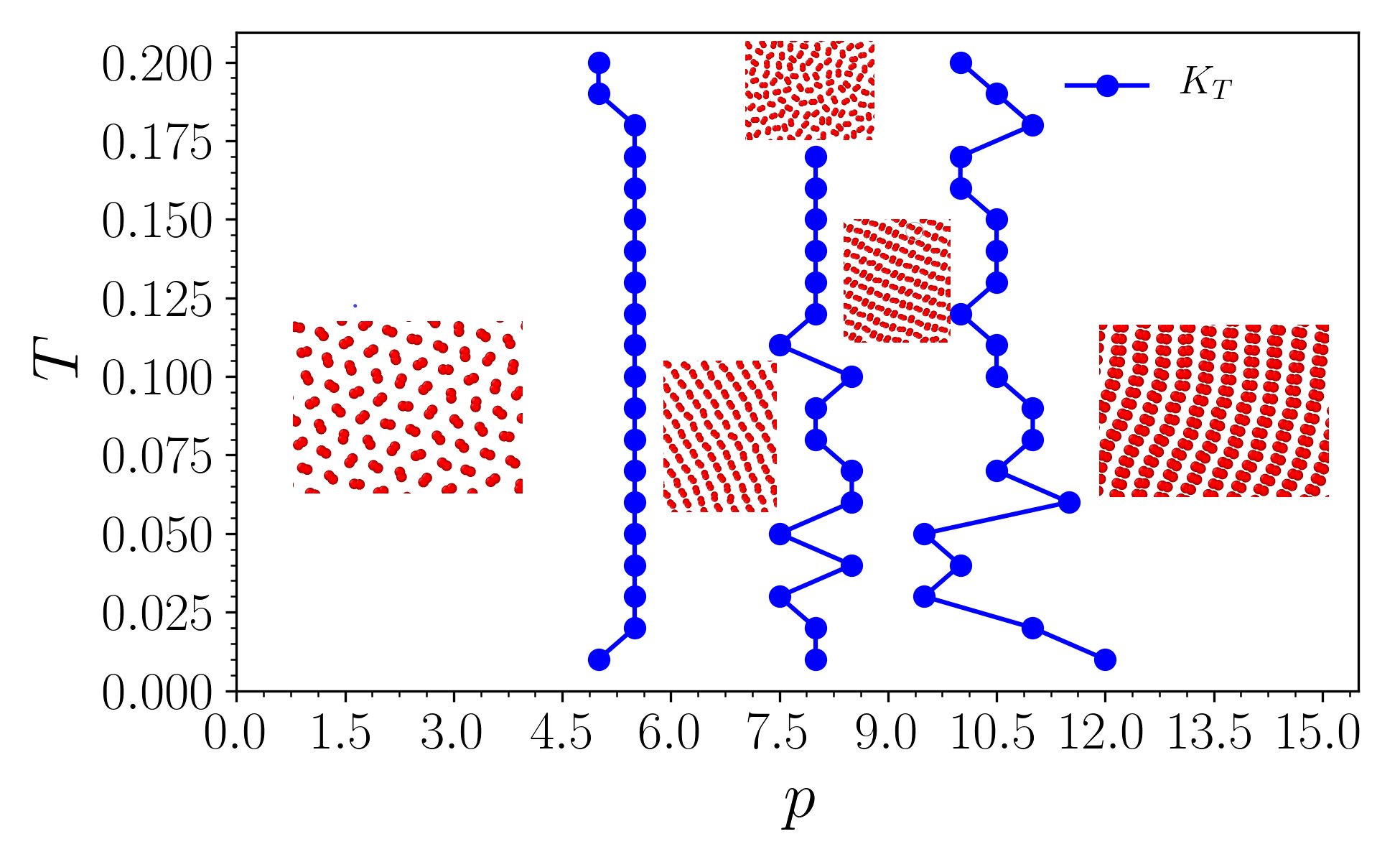}
    \caption{Phase diagram $T \times p$ obtained for the $\lambda = 0.50$. Insets: the systems snapshots.}
    \label{fig: phase-diagram-lamb05}
\end{figure*}

Thus, we present the $T \times p$ phase diagram in Fig.~\ref{fig: phase-diagram-lamb05}. At low pressures, the system assumes a structure where the center of mass are located in a triangular phase - the low density triangular (LDT) struture. At intermediate values of the simulated pressures, $p\approx 5.0$, the first stripe phases arises. It is characterized by the colloids assuming a end-to-end alignment -- the end-to-end stripe (EES) pattern. Compressing the system, some colloids flip and a T-like stripe (TS) pattern is observed, and for high pressures all colloids have rotated, and a side by side stripe (SSS) alignment is observed - the distinct structural patterns are depicted in the snapshots insets in Fig.~\ref{fig: phase-diagram-lamb05}.  

\begin{figure*}[ht]
    \centering
    \subfigure[]{\includegraphics[width=0.45\textwidth]{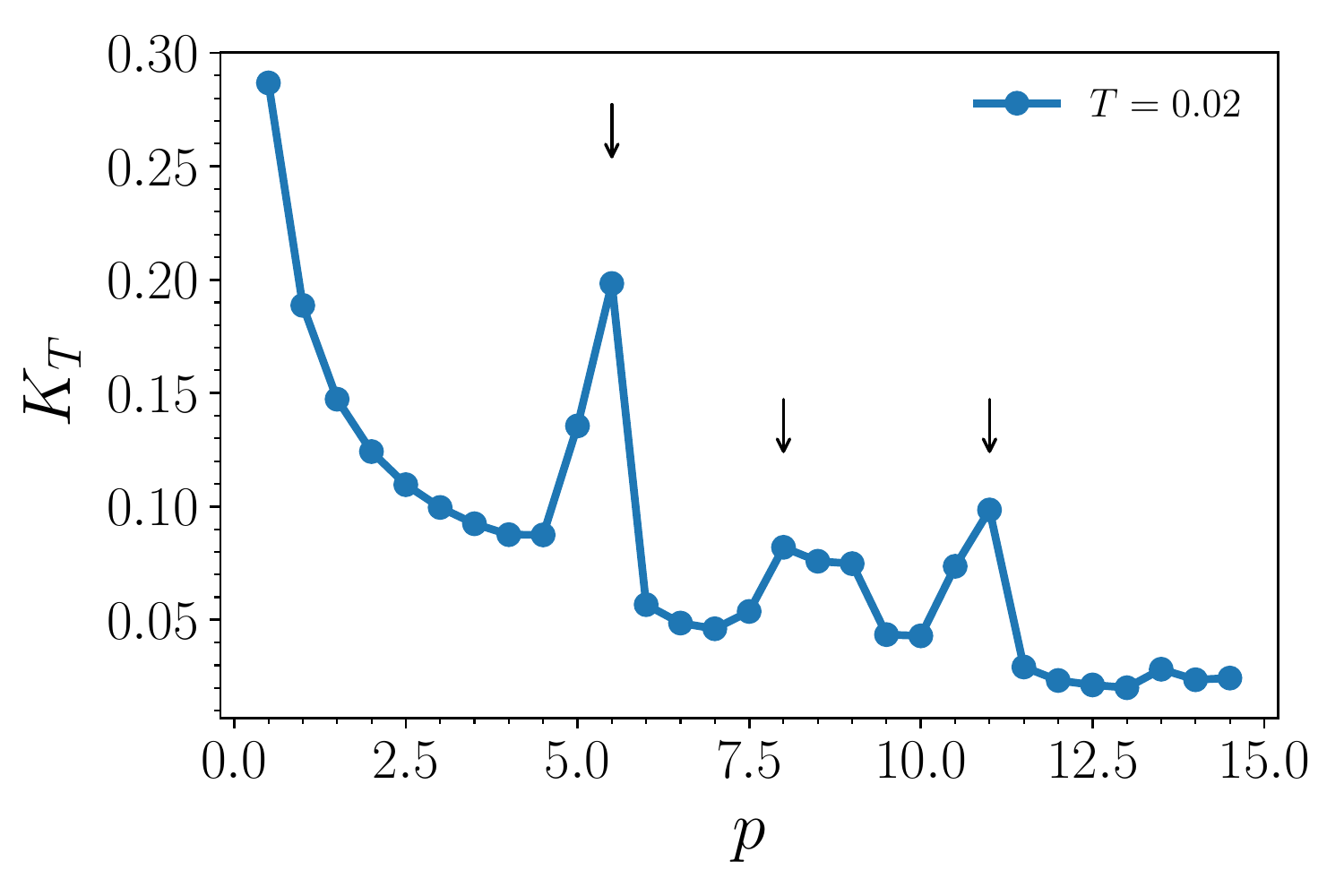}}
    \subfigure[]{\includegraphics[width=0.45\textwidth]{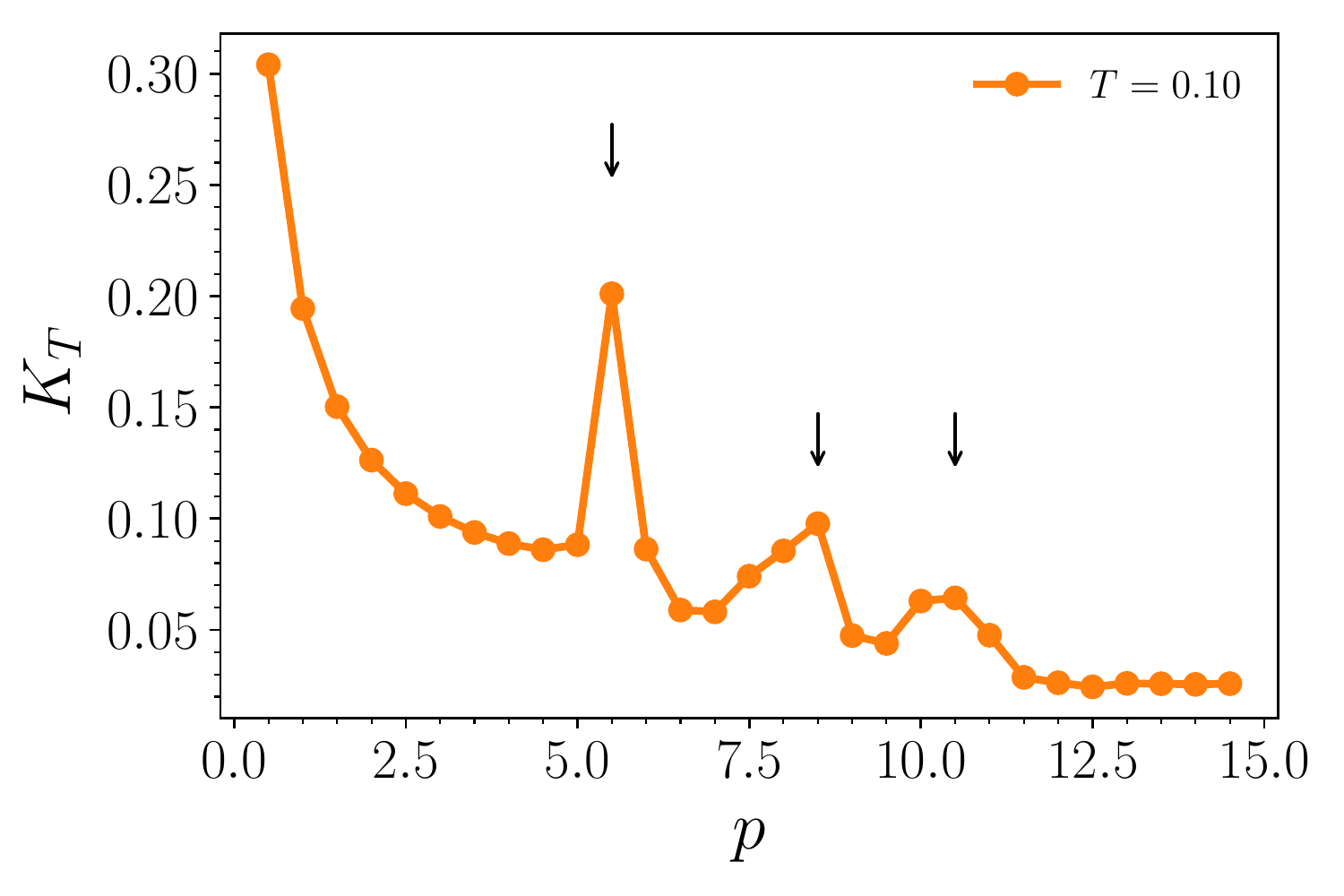}}
    \subfigure[]{\includegraphics[width=0.45\textwidth]{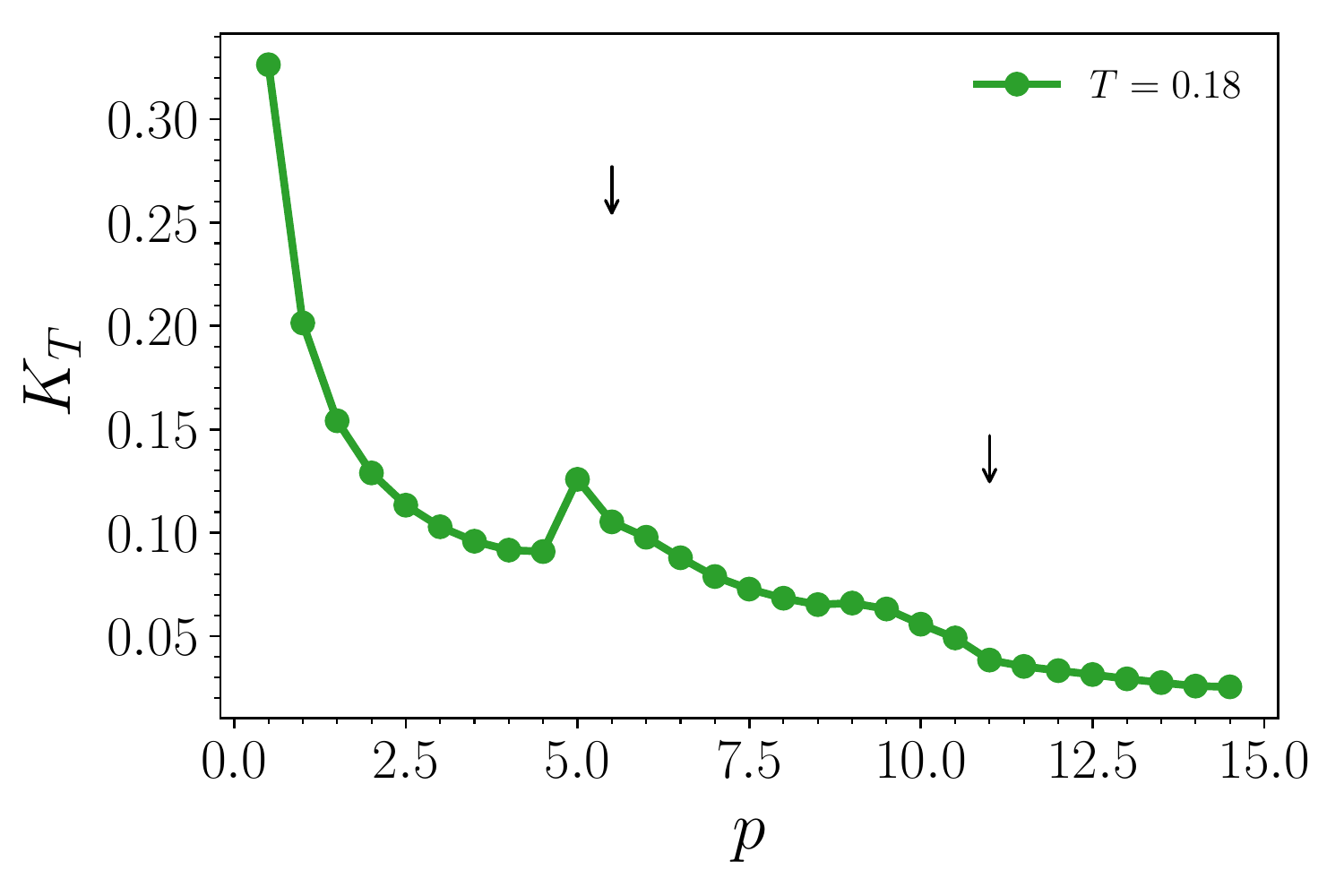}}
    \subfigure[]{\includegraphics[width=0.45\textwidth]{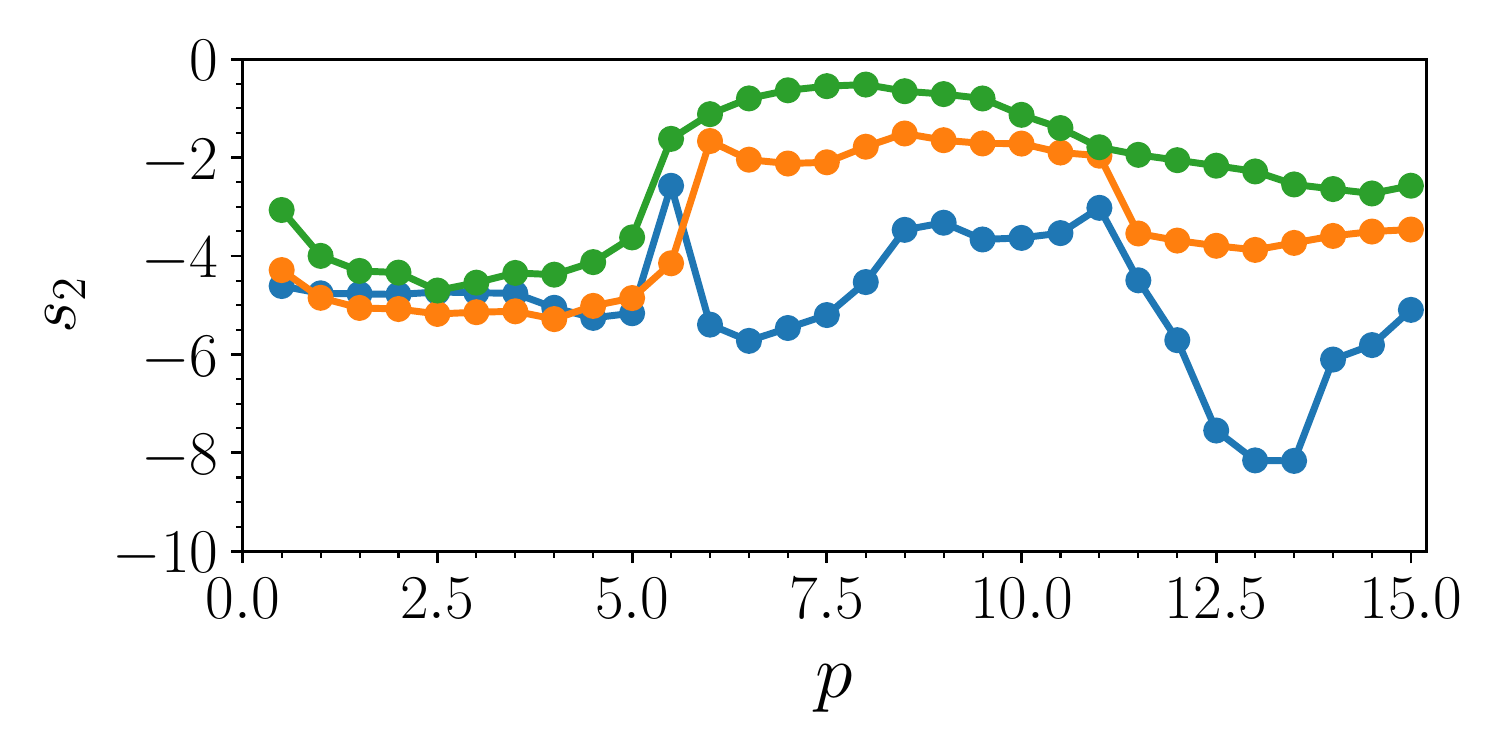}}
    \caption{The Isothermal compressibility $K_T$ as function of the system's pressure $p$ for (a) $T = 0.02$, (b) $T = 0.10$, (c) $T = 0.18$ and for these same isotherms (d) the excess entropy $s_2$ as function of the system's pressure $p$. From (a) to (c) arrows indicates a maxima or a discontinuity.}
    \label{fig: kt+s2-lamb05}
\end{figure*}

In order to identify the location of the transition between the distinct patterns we analyzed the isothermal compressibility, $K_T$. The analysis of $K_T$ is a useful tool to observe conformational changes in the system structure by exhibiting a discontinuity or a maxima in its curve. Then, their maxima are indicated by the blue lines in the phase diagram, Fig.~\ref{fig: phase-diagram-lamb05}. The $K_T$ dependence with $p$ is show in Fig.~\ref{fig: kt+s2-lamb05} (a) to (c) for low, intermediate and high temperature isotherms, namely $T = 0.02$, $0.10$, and $0.18$ respectively. The first conformational change is indicated by the maxima at $p = 5.5$ for the low temperature and $p = 5.0$ for both intermediate and high temperature isotherms - the first arrow in Fig.~\ref{fig: kt+s2-lamb05} (a) to (c). However, distinct structures are observed before and after this maxima. The low and intermediate temperatures have a change from a triangular structure to a end-to-end stripes, as the snapshots in  Fig.~\ref{fig: phase-diagram-lamb05} show, but at higher temperatures a nematic anisotropic fluid (NAL) with polymer-like pattern is observed. As we increase pressure, we observed another maxima for the low and intermediate temperatures, correspondent to the change from end-to-end stripes to Tstripes (see second arrow in Fig.~\ref{fig: kt+s2-lamb05} (a) and (b)). However, this was not observed for the polymer-like patterns at the highest temperatures, as shown in Fig.~\ref{fig: kt+s2-lamb05} (c) for $T = 0.18$. However, compressing the system all temperatures show a transition to the side by side arrangement, as the maxima (see third arrow in Fig.~\ref{fig: kt+s2-lamb05} (a) and (b)) or a discontinuity (see second arrow in Fig.~\ref{fig: kt+s2-lamb05} (c)) indicating a transition from Tstripes to side-by-side stripes for $T < 0.17$ and from the NAL phase to side-by-side stripes for $T \geq 0.17$.

For most materials, we expect a increase in the ordering under compression. A way to measure such ordering is using the pair excess entropy\cite{Lascaris2010,Sharma2006,Errington2001}. It is shown in Fig.~\ref{fig: kt+s2-lamb05} (d) for the isotherms $T = 0.02$, $0.10$ and $0.18$. We observed that for $p = 5.5$ for the lower temperatures and $p = 5.0$ $s_2$ increases, meaning that the system is likely changing from ordered structures to disordered structures. As we mapped by the means of $K_T$ and from the systems snapshots, the molecules are changing at this pressure from triangular structures to end-to-end stripes (for low temperatures) and to polymer-like structures (at high temperatures). Although unexpected, this increase in $s_2$, or increase in the disorder, under compression was reported for the monomeric case of this HCSC model in 3D and 2D~\cite{BarrosDeOliveira2006,Bordin2018, Cardoso2021} - in fact, this is the so-called structural anomaly and can be pointed out by an increase in the pair excess entropy $s_2$ given by Eq.~\ref{eq: s2} as the system is compressed. 

Nonetheless, for the low temperatures we can observe that after this first maxima $s_2$ tends to decrease showing that end-to-end stripes is gaining ordering as we increase pressure up to the point that it starts to change its conformation again to Tstripes structures depicting another maxima at $p = 8.0$. The later behavior is observed again until pressures reaches the point where packing is so high that the structure changes from Tstripes to side-by-side stripes for $p > 10.0$. For the high temperatures, where we found the fluid phase, $s_2$ tends to decrease from the first maxima as we increase pressure. This means that packing induced by compression enforces the system to order itself also in a side-by-side structure for $p > 10.0$.

\begin{figure*}[ht]
    \centering
    \subfigure[]{\includegraphics[width=0.45\textwidth]{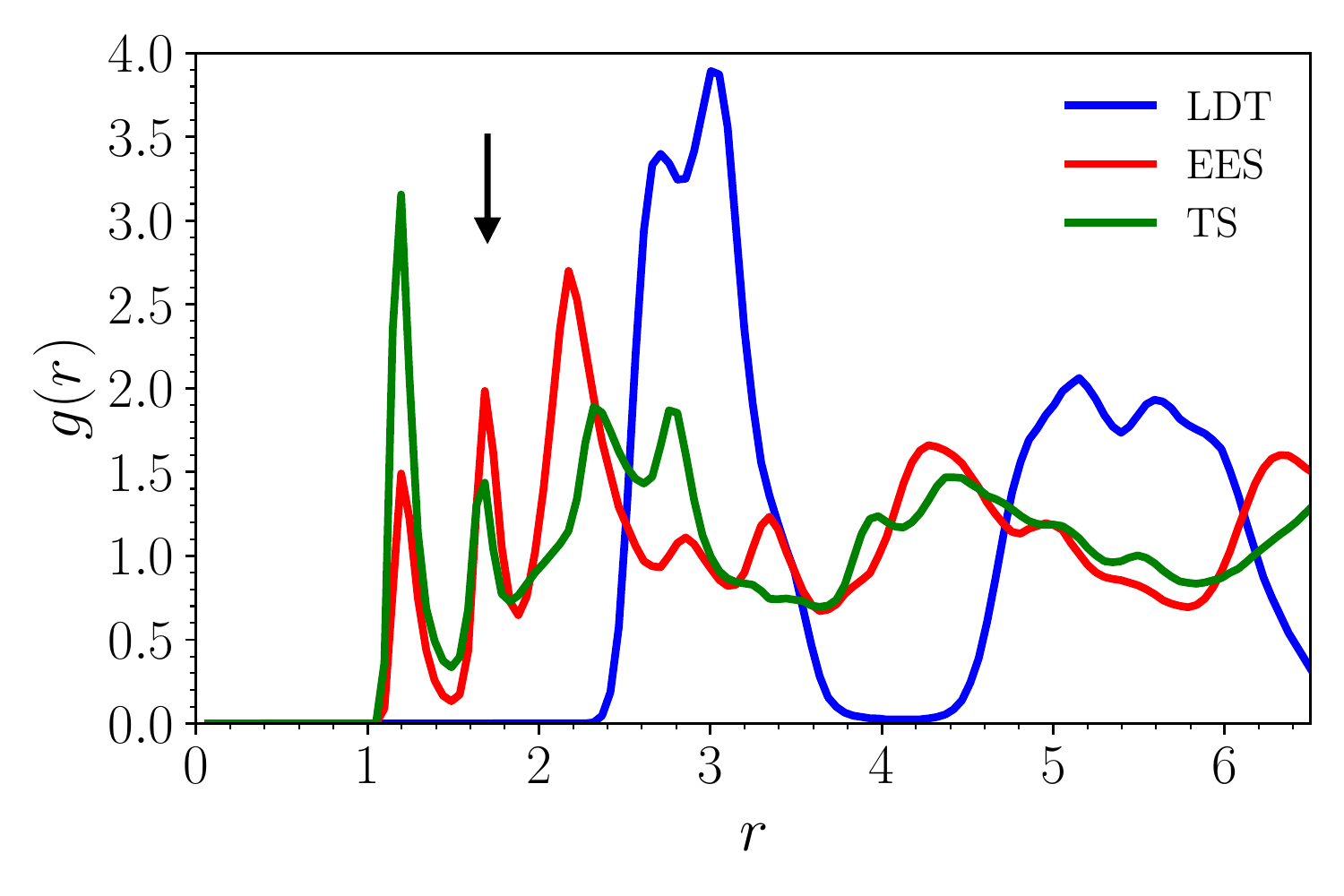}}
    \subfigure[]{\includegraphics[width=0.45\textwidth]{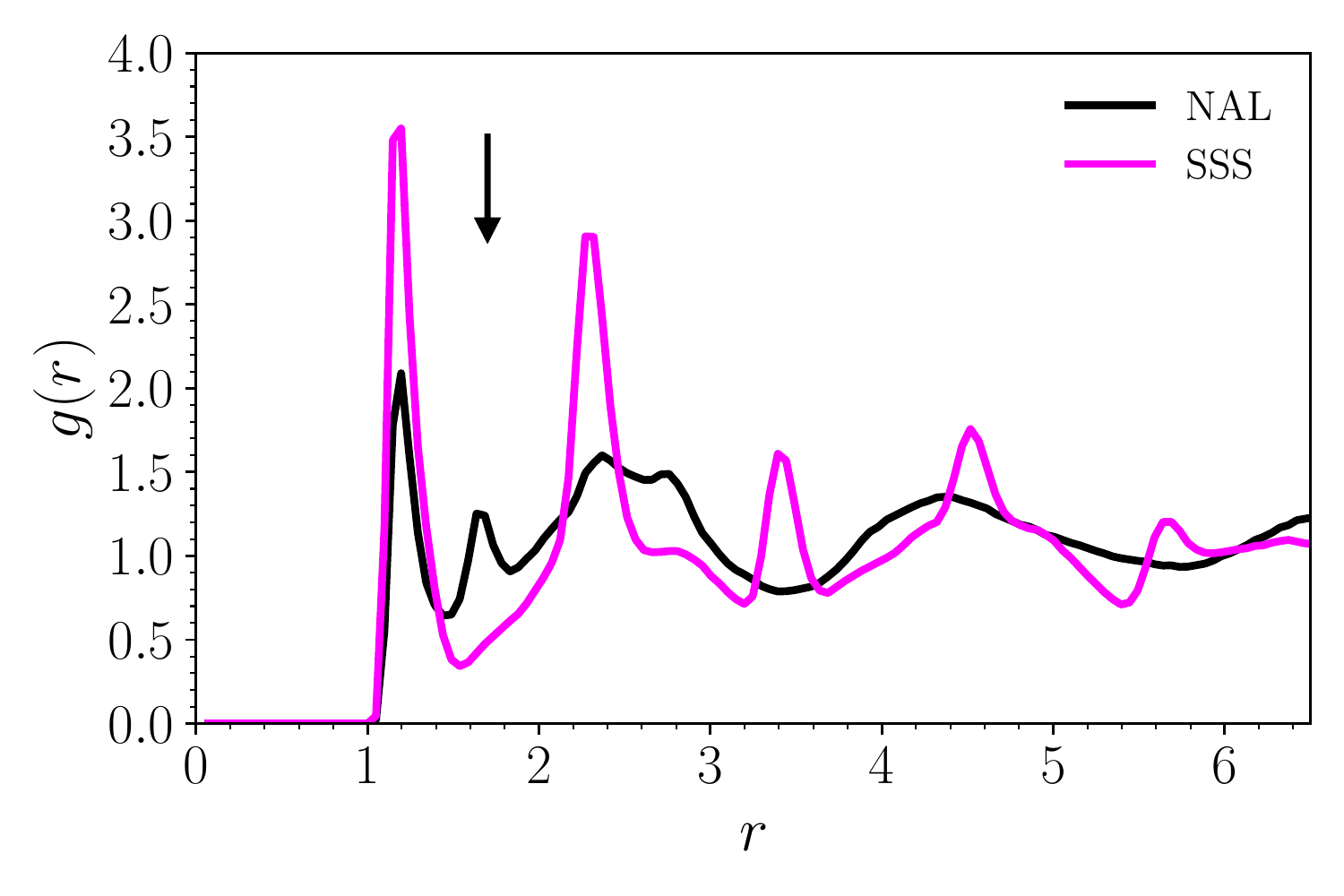}}
    \caption{The RDF of all the structures observed as we increase the system's compression. (a) Are the RDF for the LDT, EES and TS structures. (b) are the RDF for the fluid and SSS structures.}
    \label{fig: rdf-lamb05}
\end{figure*}

As usual for HCSC systems, the structural behavior can be understood by the competition between induced by the presence of two length scales in the potential interaction~\cite{BarrosDeOliveira2006,jagla1999b}. Here we have another ingredient: the dumbbell anisitropy acting as an extra length scale~\cite{nogueira2022patterns}. Is this sense, at low pressures the first coordination shell in HCSC colloid are located at the second length scale. The packing effect induced by the system's compression is then observed by the movement of the coordination shells towards the first length-scale $r_1$, and can be pointed out by the radial distribution function RDF. We illustrate the RDF in Fig.~\ref{fig: rdf-lamb05} (a) and (b) for all the different structures observed as we compress the system from $p = 0.5$ to $p = 15.0$.

As mentioned few lines ago, the first interesting structural change that occurs is the movement of the colloids first coordination shells from furthest to nearest length scale. It can be observed by comparing the RDFS for LDT to EES and TS, shown in Fig.~\ref{fig: rdf-lamb05} (a). In agreement with our previous results~\cite{nogueira2022patterns}, we observe that the anisotropic parameter $\lambda$ -- or the intramolecular rigid distance -- leaded to a new characteristic length scale at a distance $r_1 + \lambda$. This point is indicated by the arrows in Fig.~\ref{fig: rdf-lamb05} (a) and (b). Here, is important to address how the occupation at this distance plays a major role in the stripe patterns. The EES shows the higher occupancy, that decreases under compression to the TS phase and vanishes in the SSS pattern. This suggests that $\lambda$ can be employed to control the stripe pattern~\cite{nogueira2022patterns}. 

\begin{figure*}[ht]
    \centering
    \subfigure[]{\includegraphics[width=0.475\textwidth]{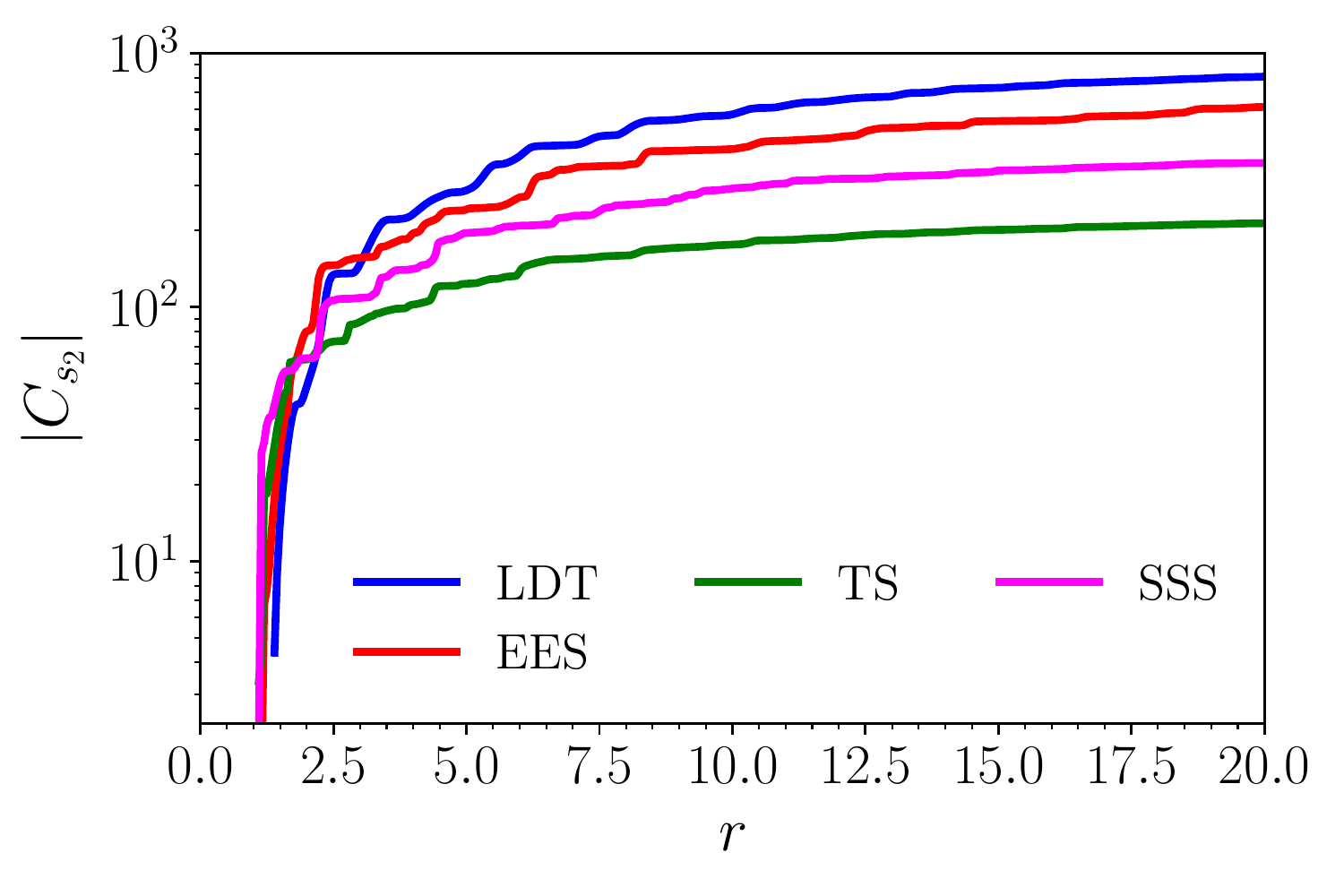}}
    \subfigure[]{\includegraphics[width=0.475\textwidth]{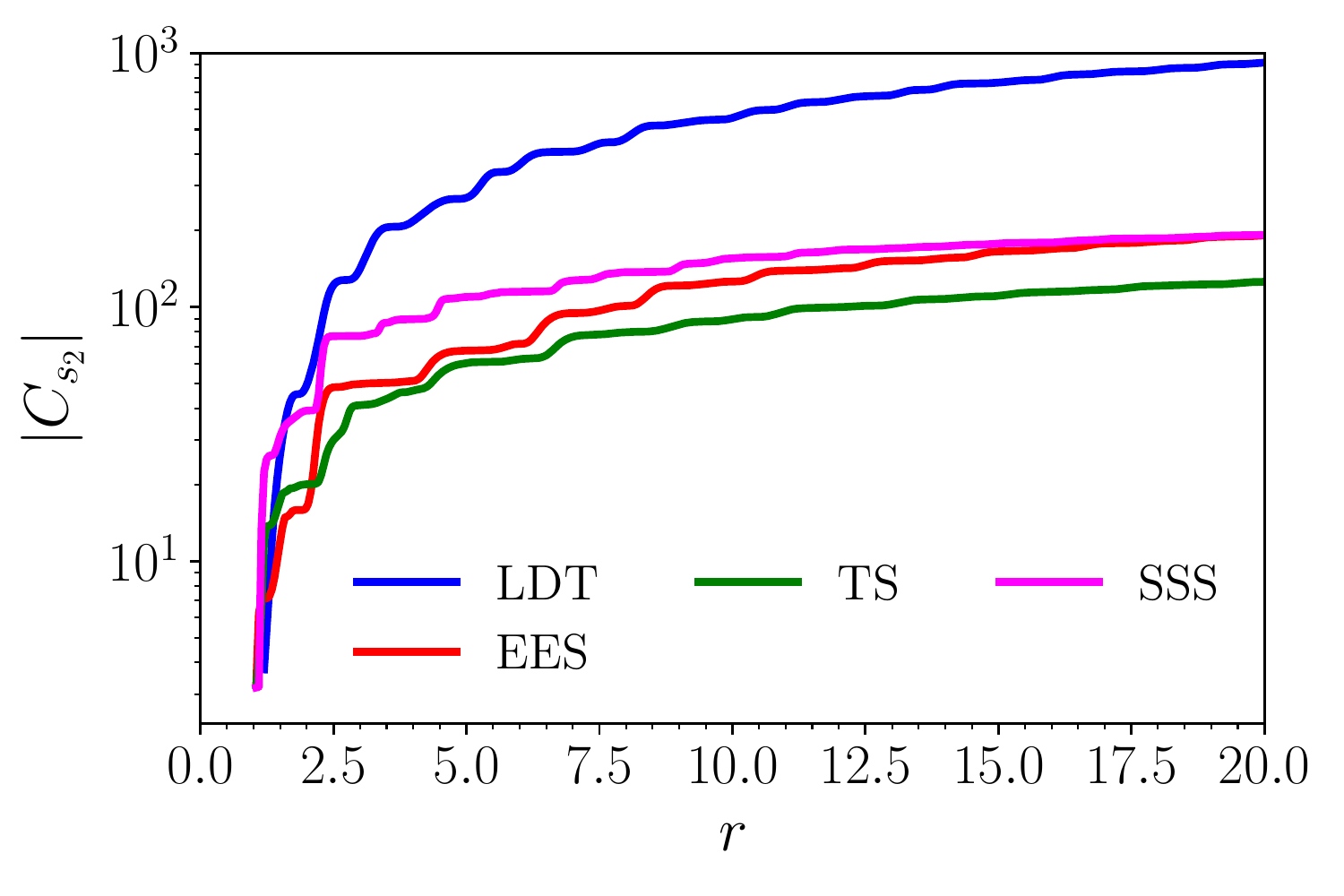}}
    \subfigure[]{\includegraphics[width=0.475\textwidth]{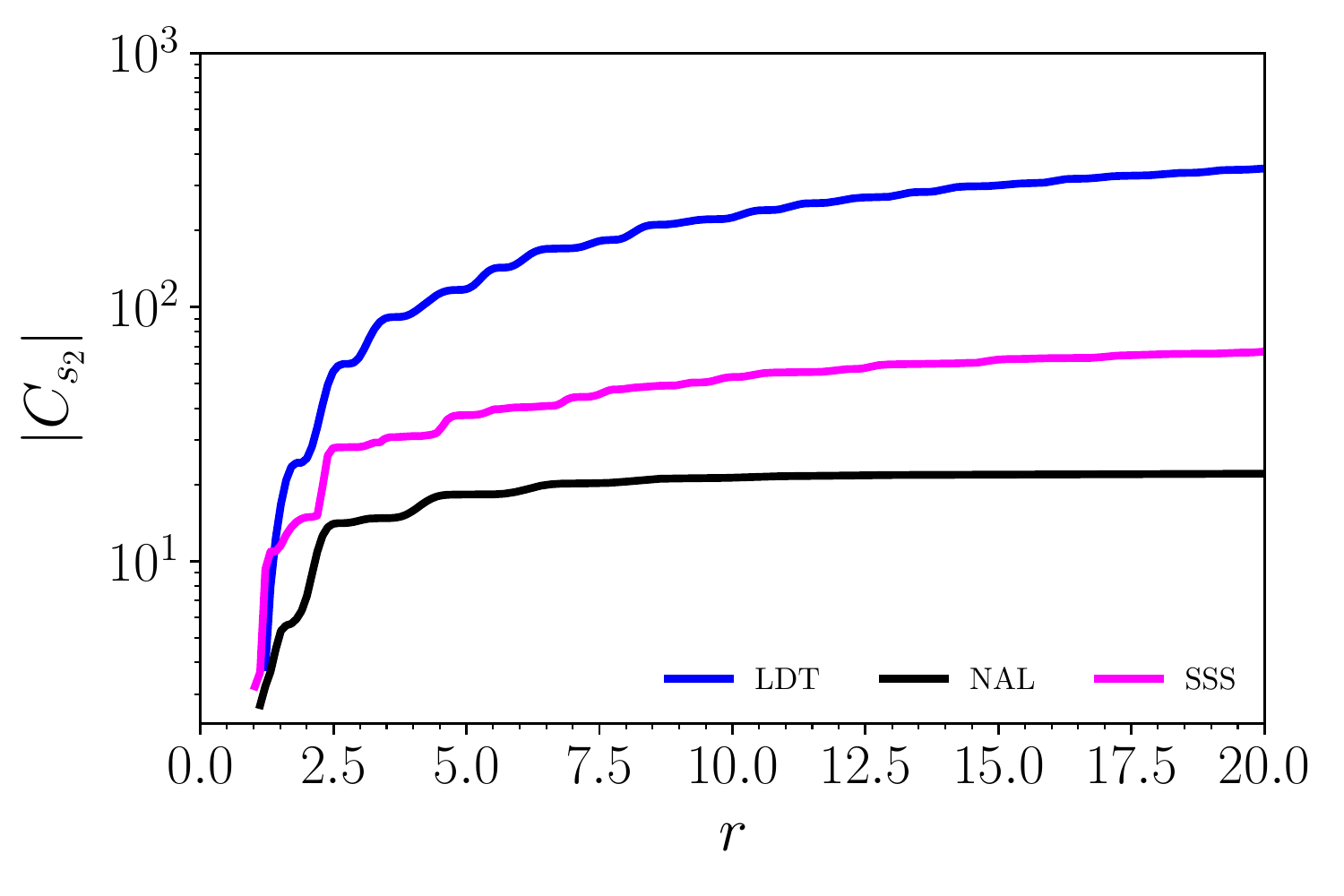}}
    \caption{(a) to (c) are the cumulative two-body excess entropy $|C_{s_2}|$ for the distinct structures observed.}
    \label{fig: cs2-lamb05}
\end{figure*}

The cumulative two-body entropy $|C_{s_2}|$ \cite{franzese2010,Marques21a} can give insights about the long range ordering. In Fig~\ref{fig: cs2-lamb05} (a) to (c) we illustrate $|C_{s_2}|$ for all the distinct phases observed for the isotherms $T = 0.02$, $0.10$ and $0.18$, respectively. As we can see in the blue curves, when the system is in the LDT phase, $|C_{s_2}|$ increases with the distance until $r < 17.5$ indicating a shorter and longer range ordering for all three isotherms. The same longer range ordering is observed in the EES structure (red curves) at low temperature. Heating this patterns, up to $T = 0.10$, for instance (see Fig.~\ref{fig: cs2-lamb05} (b)), the EES structure depicts short and intermediate ordering, reaching a plateau at $r < 15.0$. This can be related to the fact that by heating the system we are introducing more kinetic energy to the colloids which permits more relaxation, then breaking the translational ordering at long distances. For the TS structures observed at low and intermediate temperatures we observed only short and intermediate ordering since $|C_{s_2}|$ reaches a plateau at longer distances (see magenta curves in Fig.~\ref{fig: cs2-lamb05} (a) and (b)). Interesting, the reentrant NAL phase observed at the highest temperatures above the SSS and TS regions. Its translational ordering, see black curve in Fig.~\ref{fig: cs2-lamb05} (c), indicates a short range ordering depicted by a $|C_{s_2}|$ increase until reaching a plateau at distances $r > 7.5$. 

\begin{figure*}[ht]
    \centering
    \subfigure[]{\includegraphics[width=0.475\textwidth]{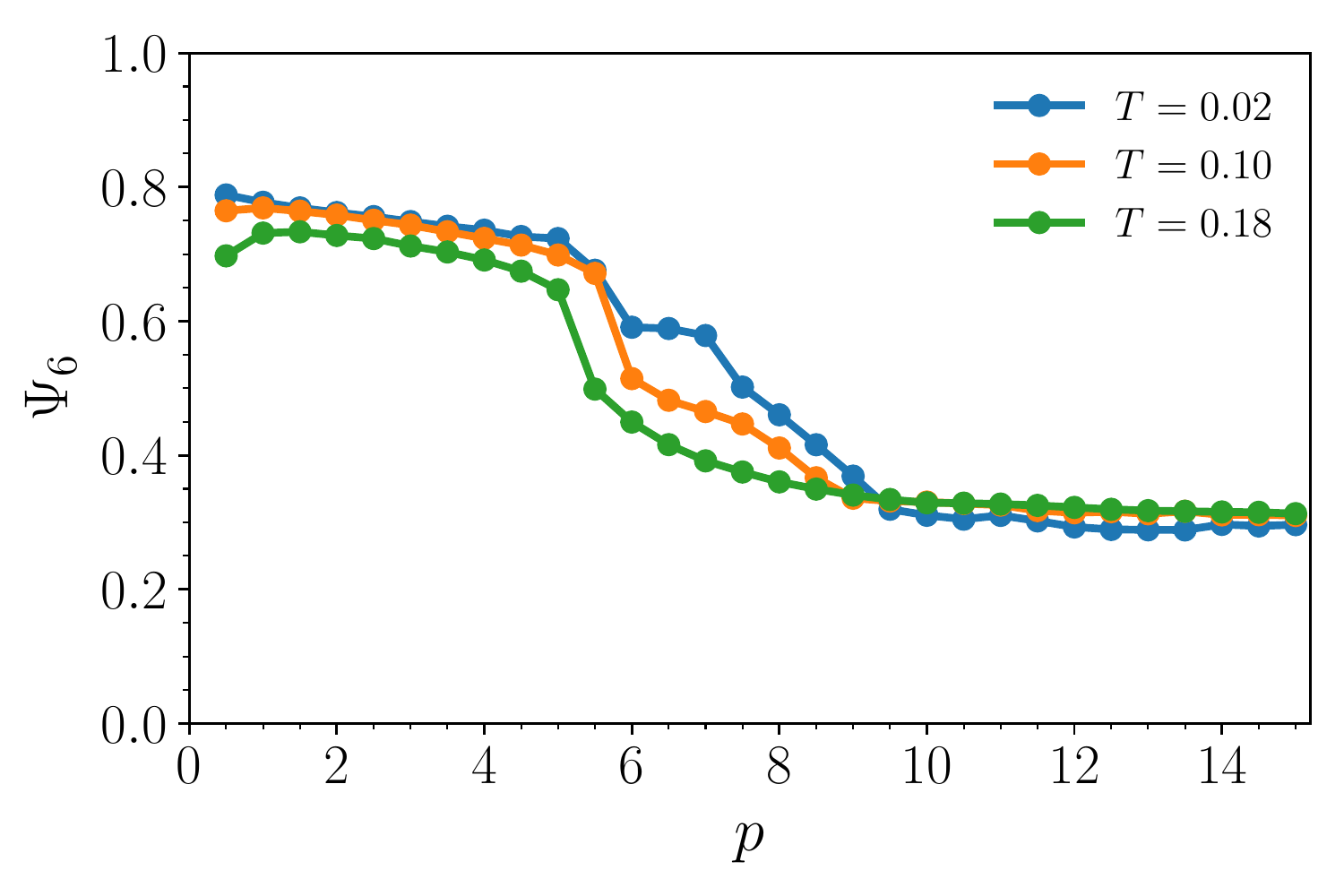}}
    \subfigure[]{\includegraphics[width=0.475\textwidth]{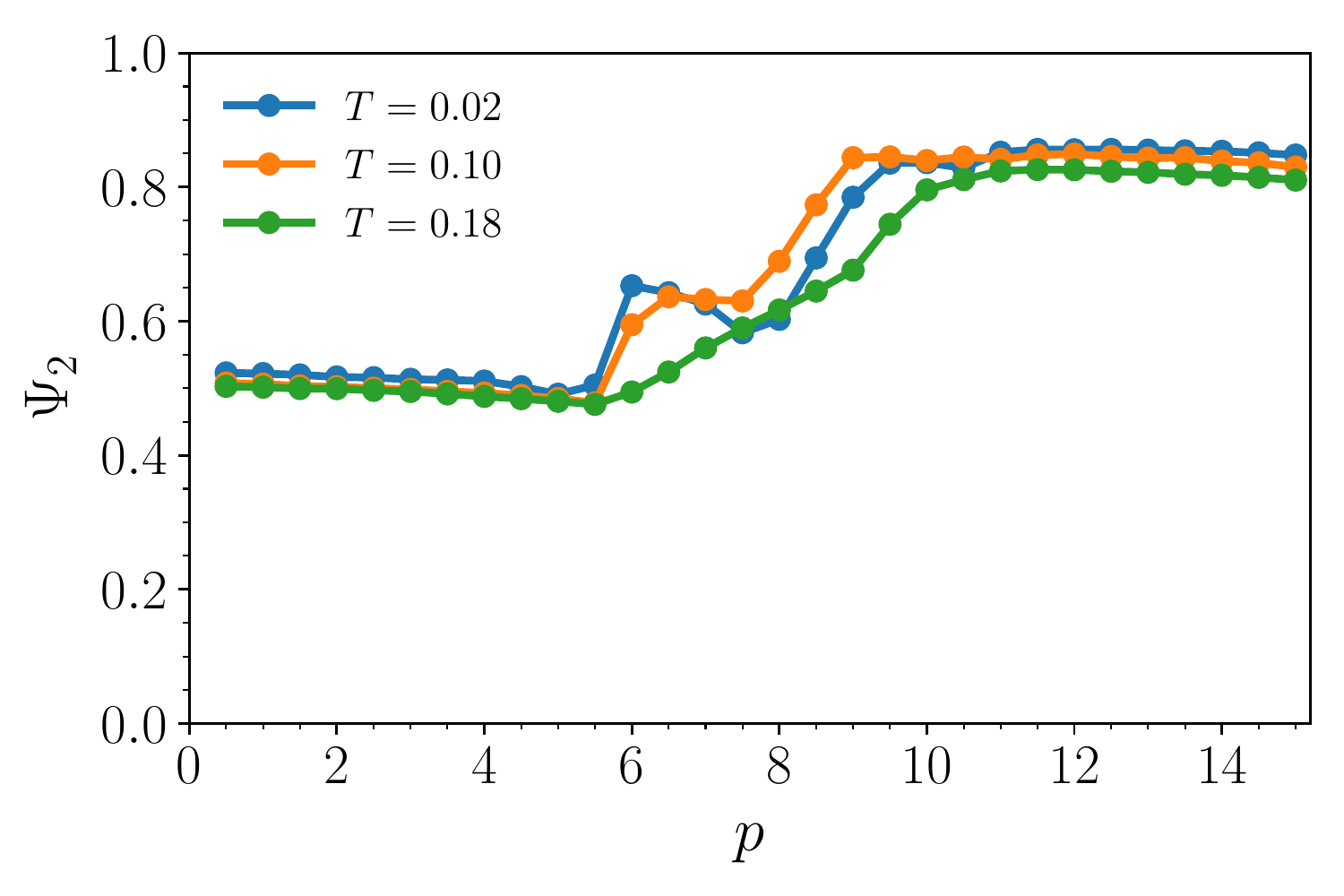}}
    \caption{(a) $\Psi_6$ and (b) $\Psi_2$ for the isotherms $T = 0.02$,$0.10$, and $0.18$ as function of the system pressure $p$.}
    \label{fig: psis62}
\end{figure*}

Usually, translational changes are followed by the orientational rearrangements. In this sense we show in Fig.~\ref{fig: psis62} the orientational bond parameters (a) $\Psi_6$ and (b) $\Psi_2$ for the isotherms $T = 0.02$,$0.10$, and $0.18$ as function of the system pressure $p$. The LDT hexagonal structure has a high $\Psi_6$. As consequence, we can observe a higher $\Psi_6$ at low pressures. This value decays as we increase the pressure and the triangular structure change to the stripes phase. In this region, the $\Psi_2$ behavior can be employed to check the transition between distinct stripes patterns. Analyzing the blue and orange curves in Fig.~\ref{fig: psis62}(b) we can see a discontinuous jump in the value of $\Psi_2$ from the pressure $p = 5.75$ to $p = 6.00$ - the LDT to EES transition. This same discontinuity is observed in the $K_T$ curve. Then, $Psi_2$ remain approximately constant until $p  = 7.5$. At this point, $\Psi_2$ increases as the dumbbells start to rotate due to the compression, which leads the ESS to TS transition. Along the TS region, $Psi_2$ keeps growing linearly with $p$ as more and more first neighbours colloids orientation change from end-to-end to T-shape - until the point where the system is in the SSS phase and $\Psi_2$ reaches a plateau. Along high temperature isotherms, as the green line in Fig.~\ref{fig: psis62}(b) for $T=0.18$, we have observed a polymer-like fluid phase. In this sense, along this isotherm we have a low pressure region with high $\Psi_6$, the LDT phase, that starts to decreases as $\Psi_2$ grows in the NAL phase until the transition to the SSS pattern at high pressures.

\begin{figure*}[ht]
    \centering
    \subfigure[]{\includegraphics[width=0.475\textwidth]{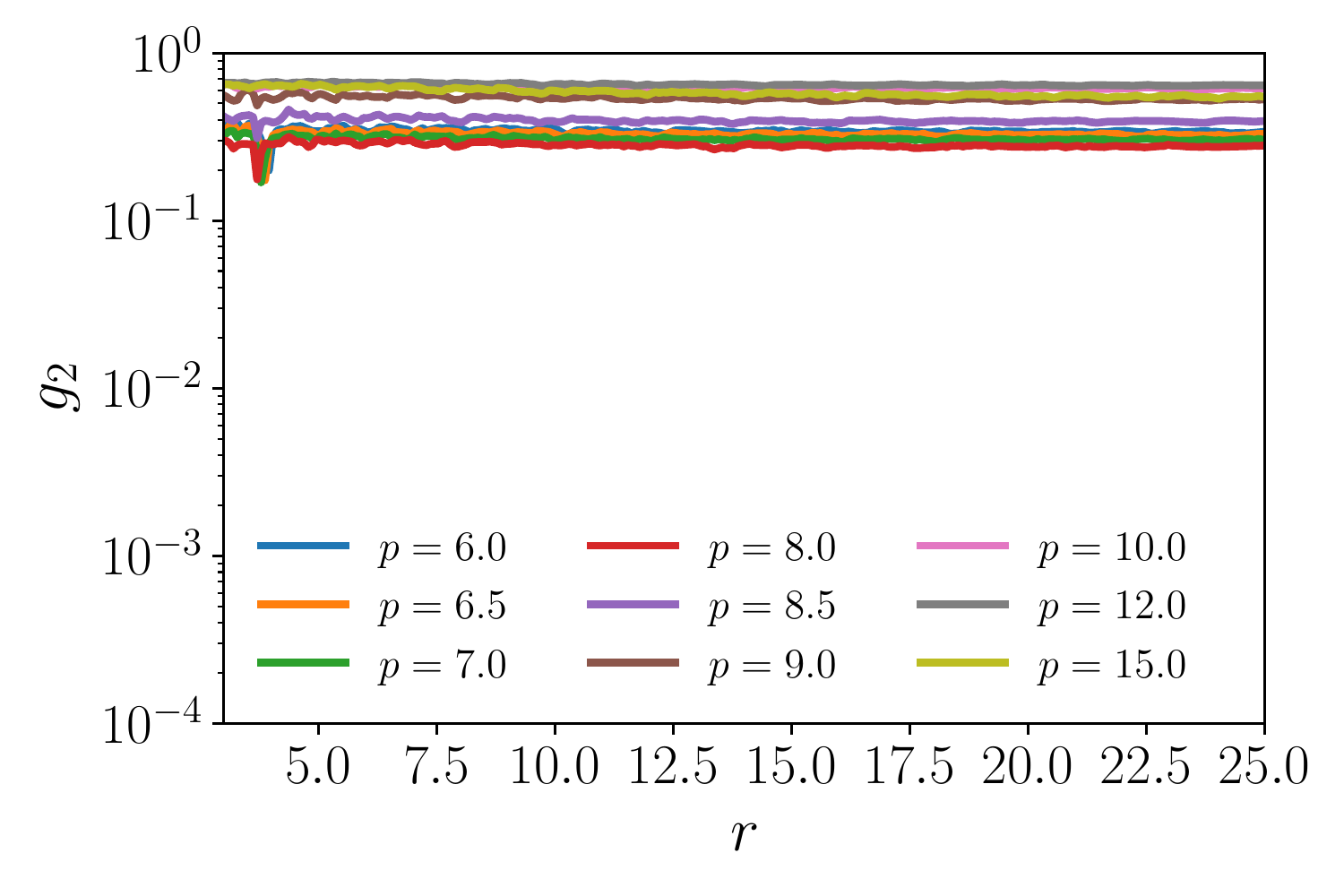}}
    \subfigure[]{\includegraphics[width=0.475\textwidth]{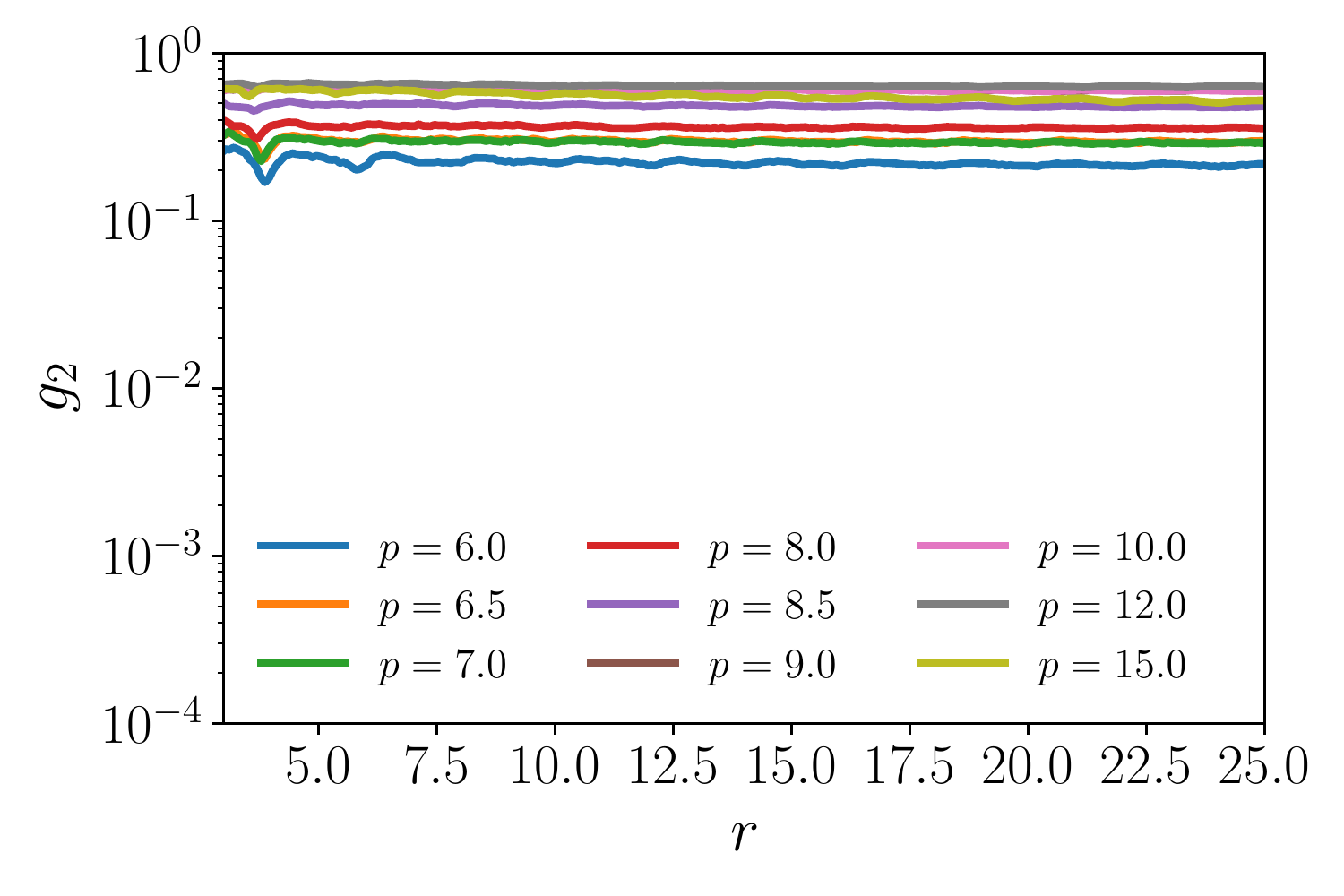}}
    \subfigure[]{\includegraphics[width=0.475\textwidth]{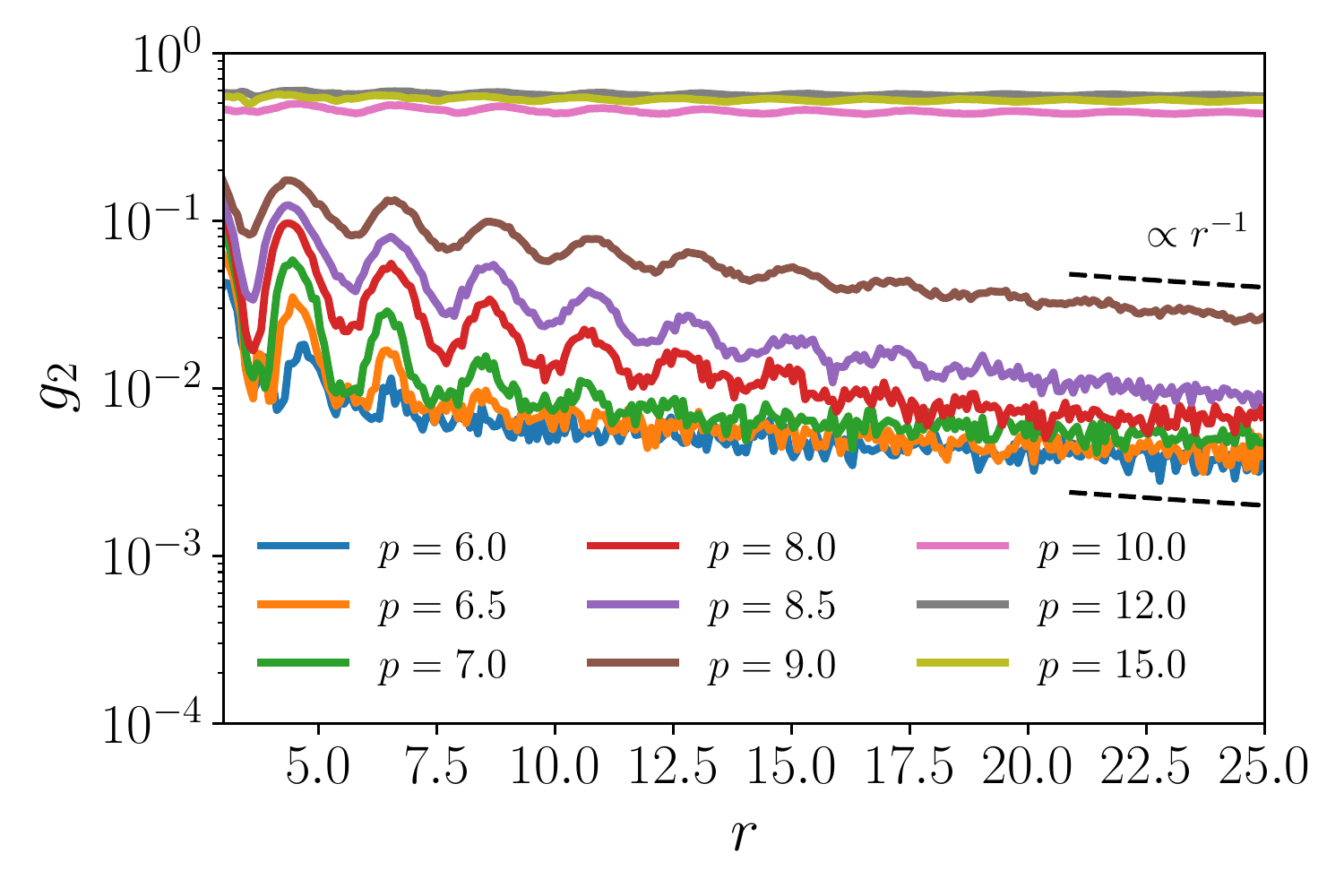}}
    \caption{
   Correlation parameter $g_2$ for (a) $T = 0.02$, (b) $0.10$ and (c) $0.18$ at distinct pressures.}
    \label{fig: g62-lamb05}
\end{figure*}

Once stripes patterns have a twofold orientational symmetry~\cite{Hurley1992}, we explore the long range orientational ordering in this region using the $g_2$ behavior, shown in Figs.~\ref{fig: g62-lamb05} (a) to (c) for the isotherms $T = 0.02$, $0.10$ and $0.18$, respectively. It is interesting to observe that even though we are compressing the system, along the low and intermediate isotherms, see Fig.~\ref{fig: g62-lamb05} (a) and (b), the $g_2$ remains approximately constant at long distances, indicating a well defined alignment of the EES, TS and SSS patterns. On the other hand, the high temperature isotherm has the polymer-like pattern phase. Along the NAL region, ranging from $p = 6.0$ to $p = 9.0$ at $T = 0.18$, $g_2$ shows a decorrelation that decays proportional to the power law $r^{-1}$ at long distances (see Fig.~\ref{fig: g62-lamb05} (c)). This power law decay is characteristic of nematic anisotropic fluidic phases~\cite{torres2022hard,kundu2021long,frenkel1985evidence,cosentino2003isotropic}. The long range orientational ordering is recovered at high pressures, in the SSS phase. 
\begin{figure*}[ht]
    \centering
    \subfigure[]{\includegraphics[width=0.475\textwidth]{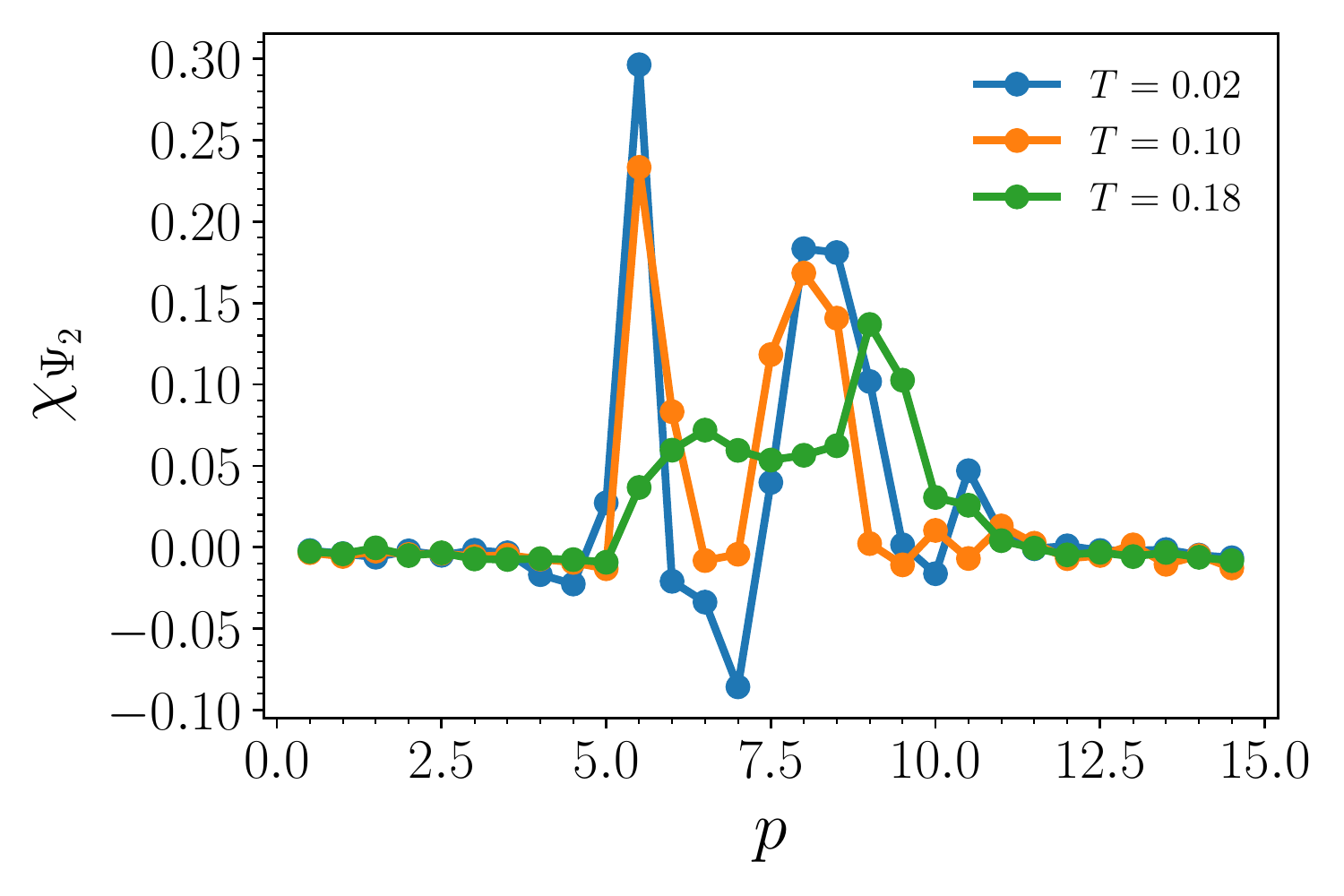}}
    \subfigure[]{\includegraphics[width=0.475\textwidth]{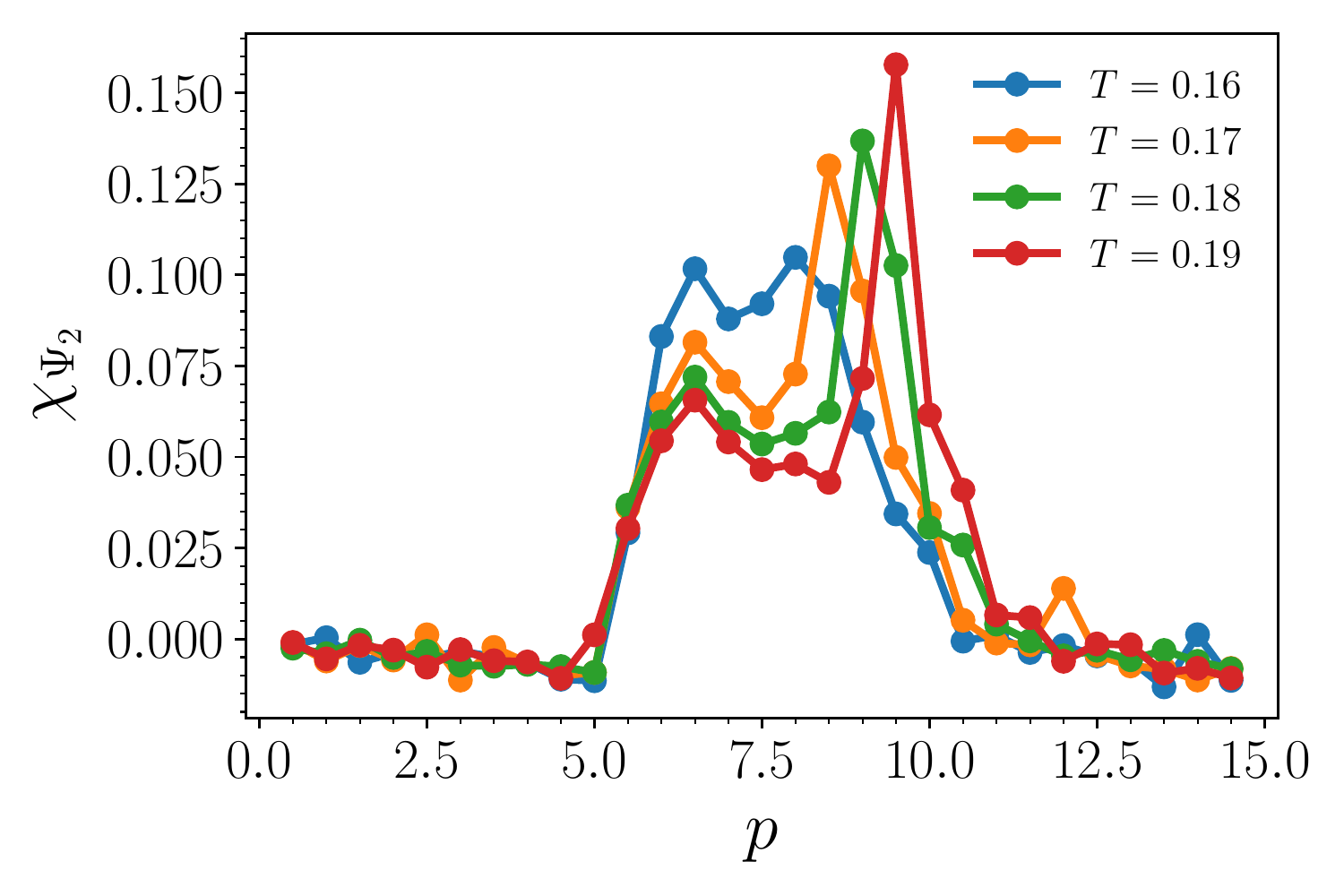}}

    \caption{(a) Pressure dependence of the isothermal bond orientational order parameter susceptibility, $\chi_{\Psi_2}$ along the same isotherms $T = 0.02, 0.10$ and 0.18 and (b) for all isotherms that cross the polymer-like fluid phase.}
    \label{fig: chipsis62}
\end{figure*}

We can also explore the transition using the isothermal bond orientational order parameter susceptibility, $\chi_{\Psi_2}$, shown in  Fig.~\ref{fig: chipsis62}(a). For the low and intermediate temperatures, the LDT to EES transition is characterized by a discontinuity in the response function $\chi_{\Psi_2}$. Then, as the system is compressed, $\chi_{\Psi_2}$ reaches a minimum close to $p = 7.25$ and increases to a maxima at $p = 7.75$. This maxima coincided with the EES to TS transition. Interesting, we observed a similar behavior in the reentrant fluid phase. Two maxima are obtained along the green curve in  Fig.~\ref{fig: psis62}(c) - and for all isotherms with fluid phase,  Fig.~\ref{fig: chipsis62}(b). This can indicate a existence of two polymer conformations in the fluid phase, one EES-like and another one TS-like. To check it, we can return to  the radial distribution function.

\begin{figure*}[ht]
    \centering
    \subfigure[]
    {\includegraphics[width=0.485\textwidth]{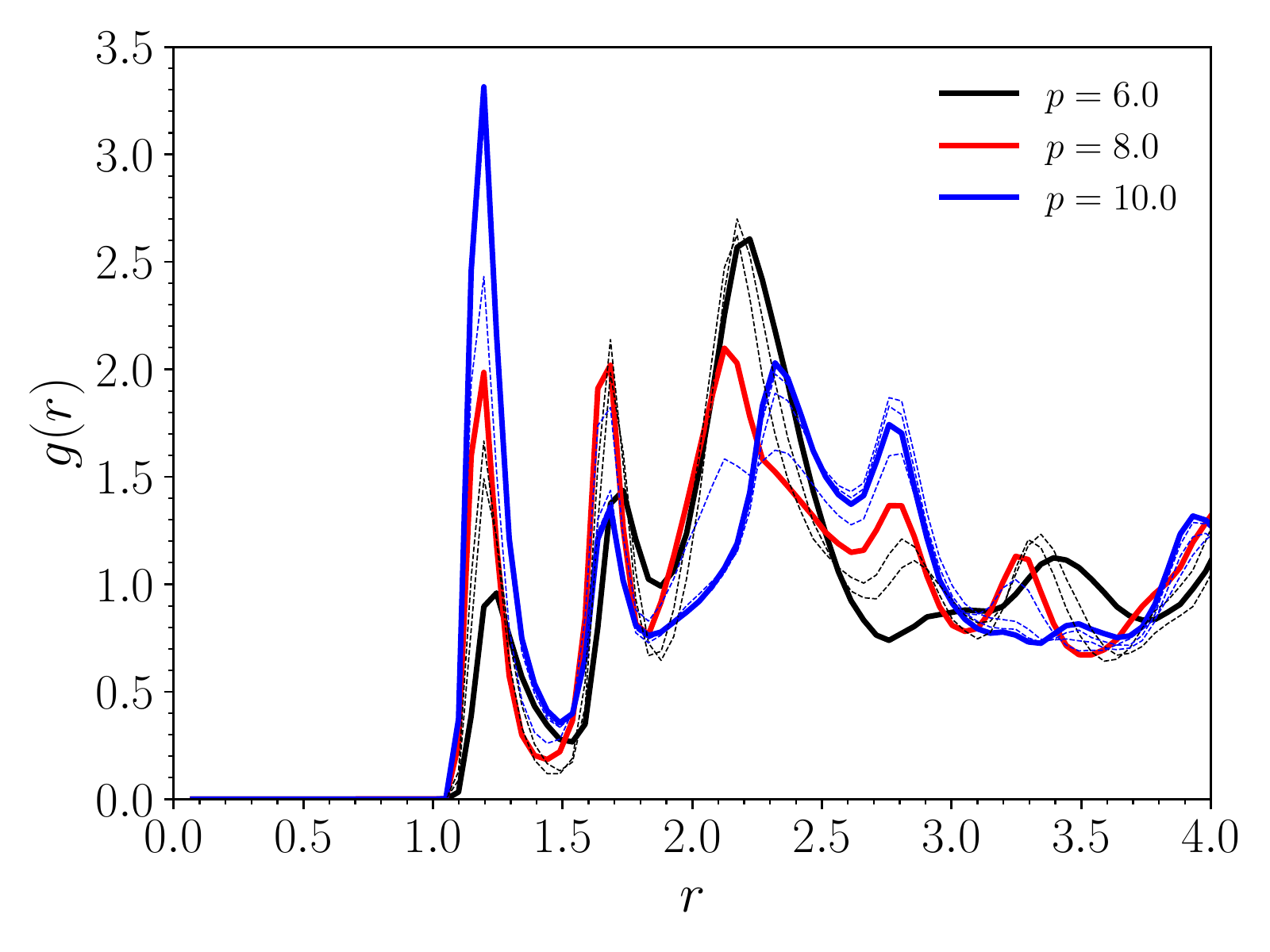}}
    \subfigure[]
    {\includegraphics[width=0.485\textwidth]{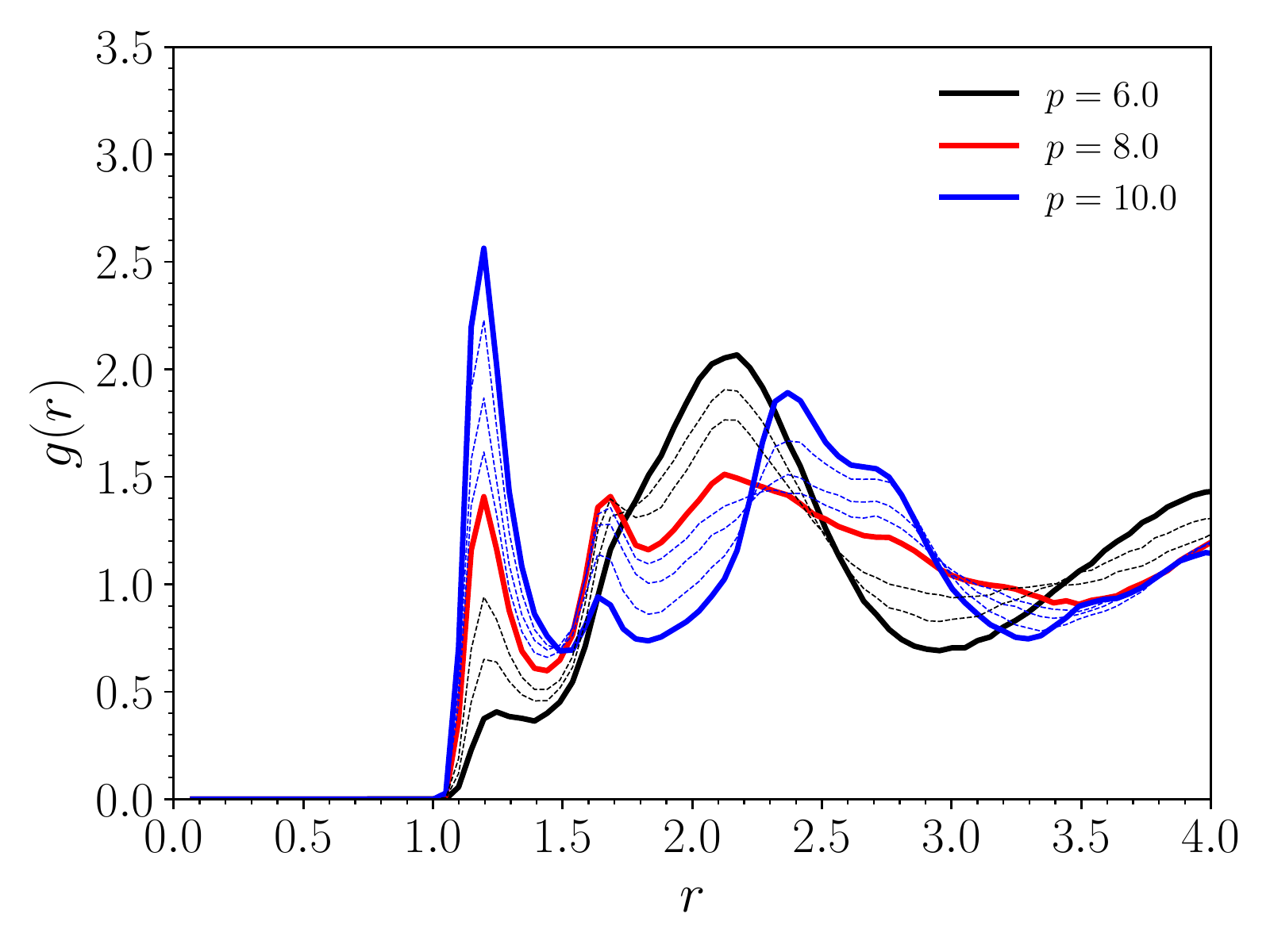}}
    \caption{RDF for pressures ranging from $p = 6.0$ to 7.50 (black curves), $p = 8.00$ (red curve) up to $p=10.0$ (blue curves) at (a) $T$ = 0.10 and (b) $T$ = 0.18.}
    \label{fig: grcompetition}
\end{figure*}

\begin{figure*}[ht]
    \centering
    \subfigure[]
    {\includegraphics[width=0.485\textwidth]{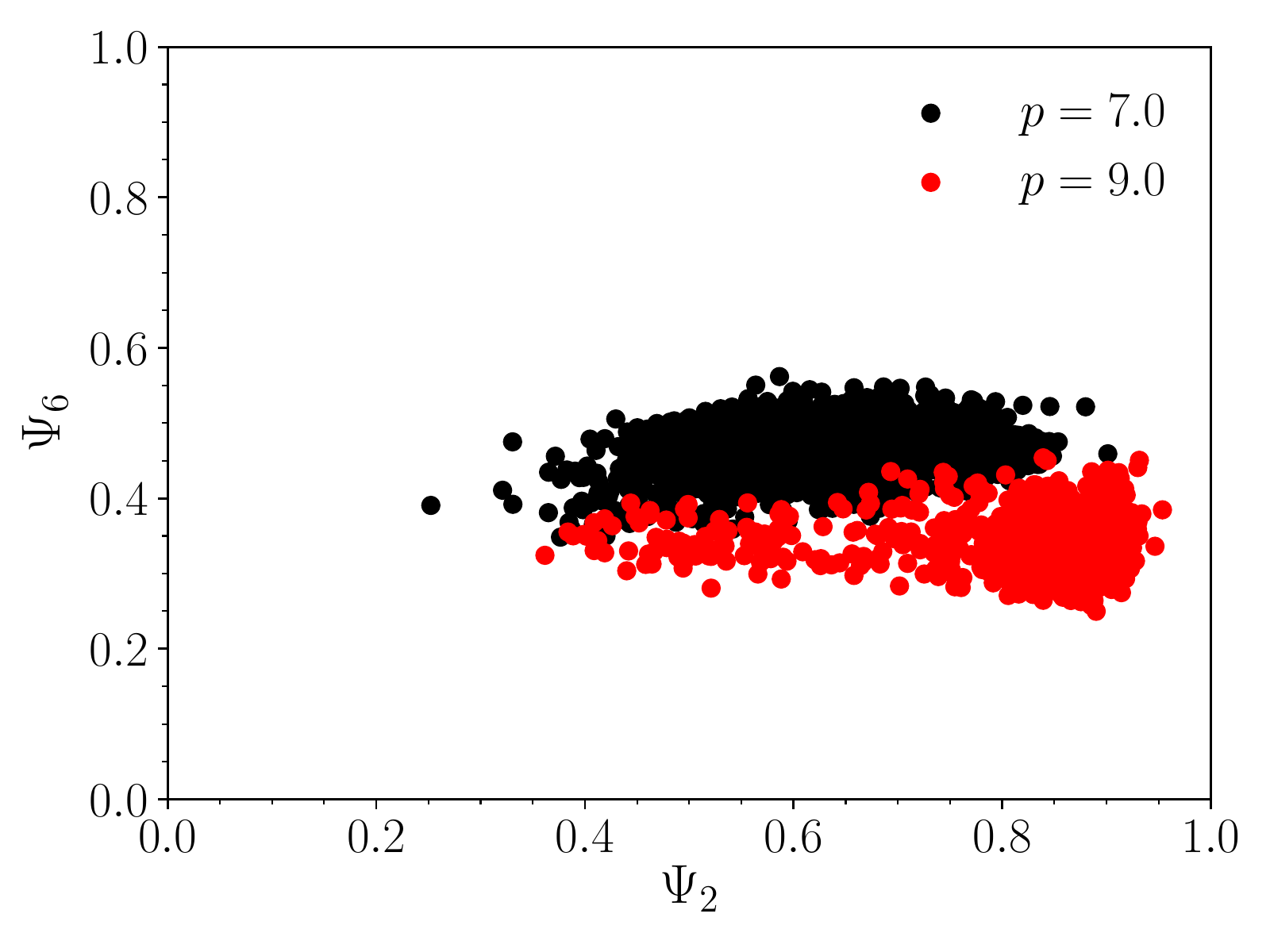}}
    \subfigure[]
    {\includegraphics[width=0.485\textwidth]{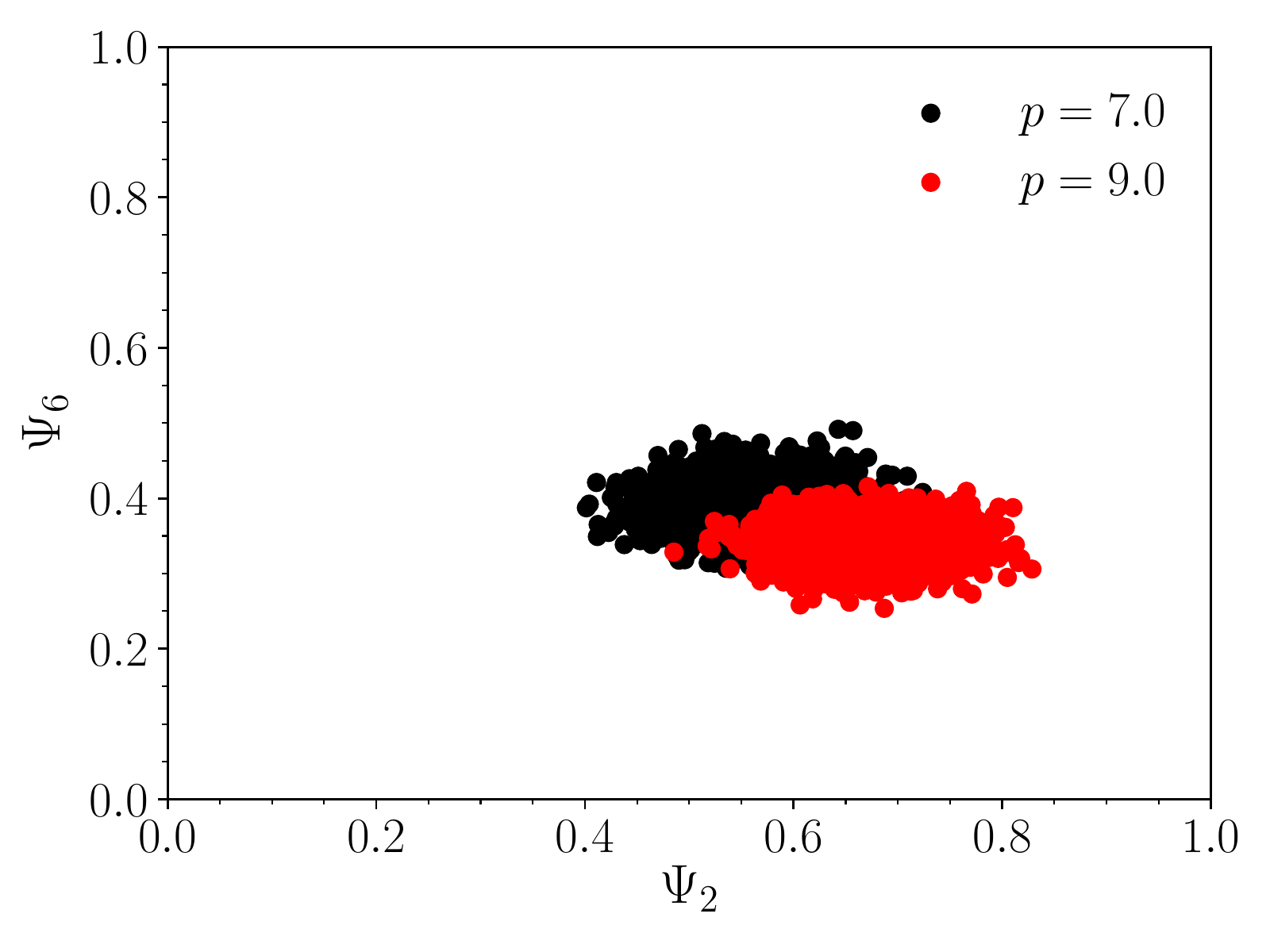}}
    \caption{(a) $\Psi_6\times\Psi_2$ scatter plots for particle sat $T = 0.10$ and two pressures: $p = 7.0$ in the EES phase and $p = 9.0$ in the TS region. (B) Same scatter plot, but at $T = 0.18$.}
    \label{fig: psicompetition}
\end{figure*}

In Fig.~\ref{fig: grcompetition}(a) we show how the structure changes under compression along the $T=0.10$ isotherm from $p = 6.00$ (thicker black line), in the EES phase, up to $p = 10.0$ (thicker blue line), in the TS phase. The intermediate pressure, $p = 8.00$, is shown by the thicker red line red. The black dotted curves goes from $p=6.25$ to $p=7.50$, and blue dotted curves for $8.50 \geq p < 10.0$. Here we can understand how the compression makes the occupancy in the characteristic length scale at $r_2 \approx 2.20$ decrease, increasing the occupancy in the hard-core length scale, $r_1 \approx 1.2$ and in the length scale induced by the dumbbell anisotropy, $r_1 +\lambda \approx 1.7$. The occupation in $r_1$ and $r_1+\lambda$ grows under compression, until $p = 8.0$, where the occupancy are approximately the same. Above this threshold, the occupancy in $r_1 +\lambda$ decreases as $p$ increases, while the peak at $r_1$ grows. It is a consequence of the EES to TS transition: the end-to-end alignment favors the higher occupation at $r_1 +\lambda$, while the TS (and SSS) patterns will favor the first characteristic length scale at $r_1$. This transition reflects in the fluid phase, as indicated in Fig.~\ref{fig: grcompetition}(b). For pressures smaller than 8.0 the occupancy in the length scale $r_2$ decreases, and the occupation in $r_1 +\lambda$ and $r_1$ rises under compression. At $p = 8.0$, the occupation in these length scales are the same, and above this pressure the peak at $r_1$ becomes higher. Not only the translational ordering indicates a competition between a end-to-end fluid and a TS-like fluid, but the orientational ordering as well. In the Fig.~\ref{fig: psicompetition}(a) we show a scatter plot of $\Psi_6\times\Psi_2$ - each point is the mean value of $\Psi_{6;2}$ for a particle at $T = 0.10$ and two pressures: $p = 7.0$ in the EES phase and $p = 9.0$ in the TS region. If we analyze the same pressures, but at $T = 0.18$, shown in a Fig.~\ref{fig: psicompetition}(b), similar values of $\Psi_6$ and $\Psi_2$ are observed, indicating a EES-like fluid phase and a TS-like fluid phase.

The competition between two structures that struggle to rule the fluid behavior can lead to anomalies. Water is the most well known anomalous fluids, with increasing evidence of a liquid-liquid phase transition (LLPT) that ends in a liquid-liquid critical point (LLCP) as a origin for their anomalies~\cite{poole1992, Gallo2016, gallo21, Verde2022}, which is connected to the phase transition observed between two glassy phases at lower temperatures~\cite{stanley1998, Stanley2000, kim20, shi20,Bachler21, Foffi21,giovambattista21,Caupin21,lucas22}. Here, although the lack of a LLPT due the purely repulsive interaction potential~\cite{Bordin2023}, our scenario seems to be similar: a competition between two conformations in the fluid phase which are connected to distinct stripes patterns. Then is natural to look for anomalous behavior.

\begin{figure*}[ht]
    \centering
    \subfigure[]
    {\includegraphics[width=0.485\textwidth]{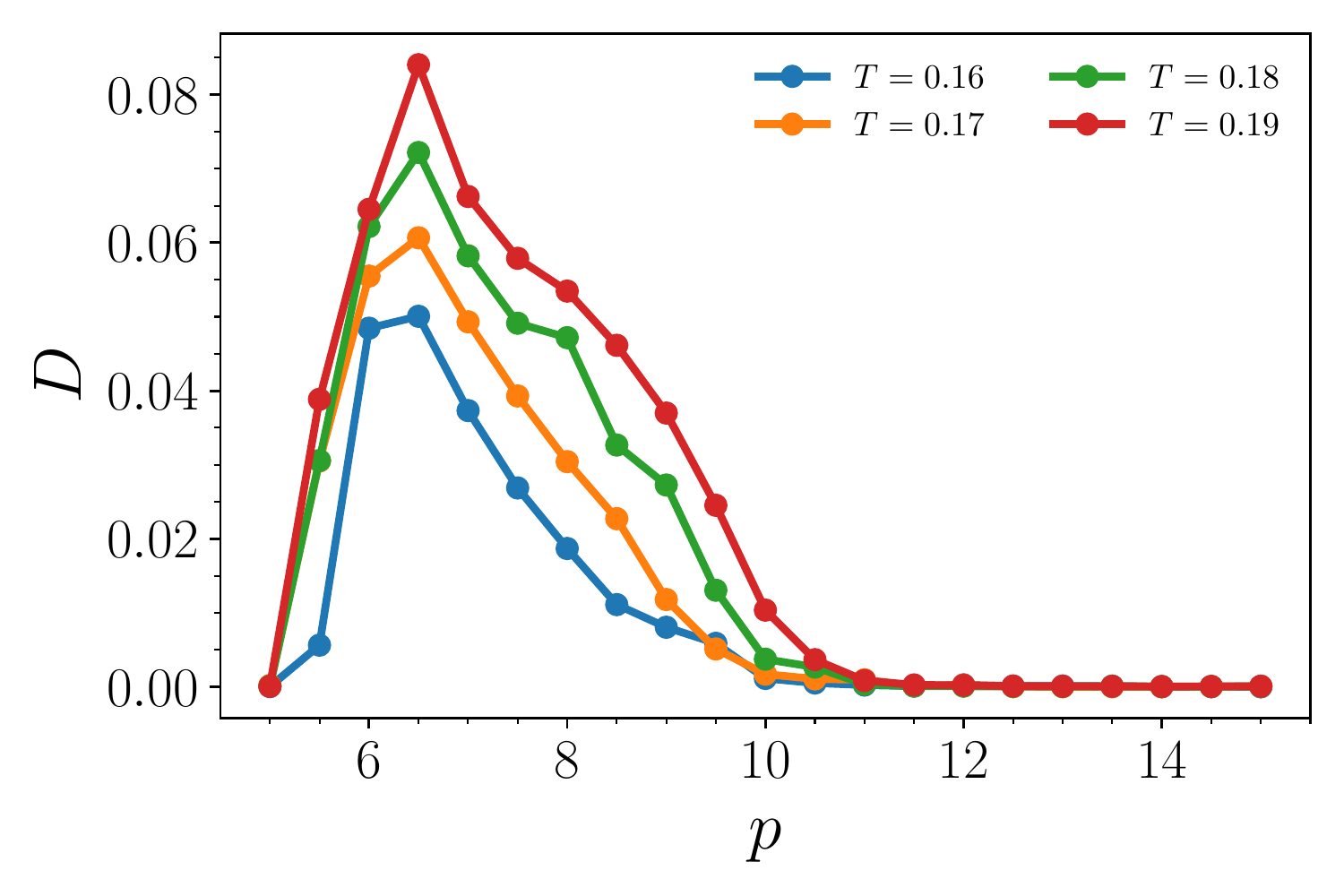}}
    \subfigure[]
    {\includegraphics[width=0.485\textwidth]{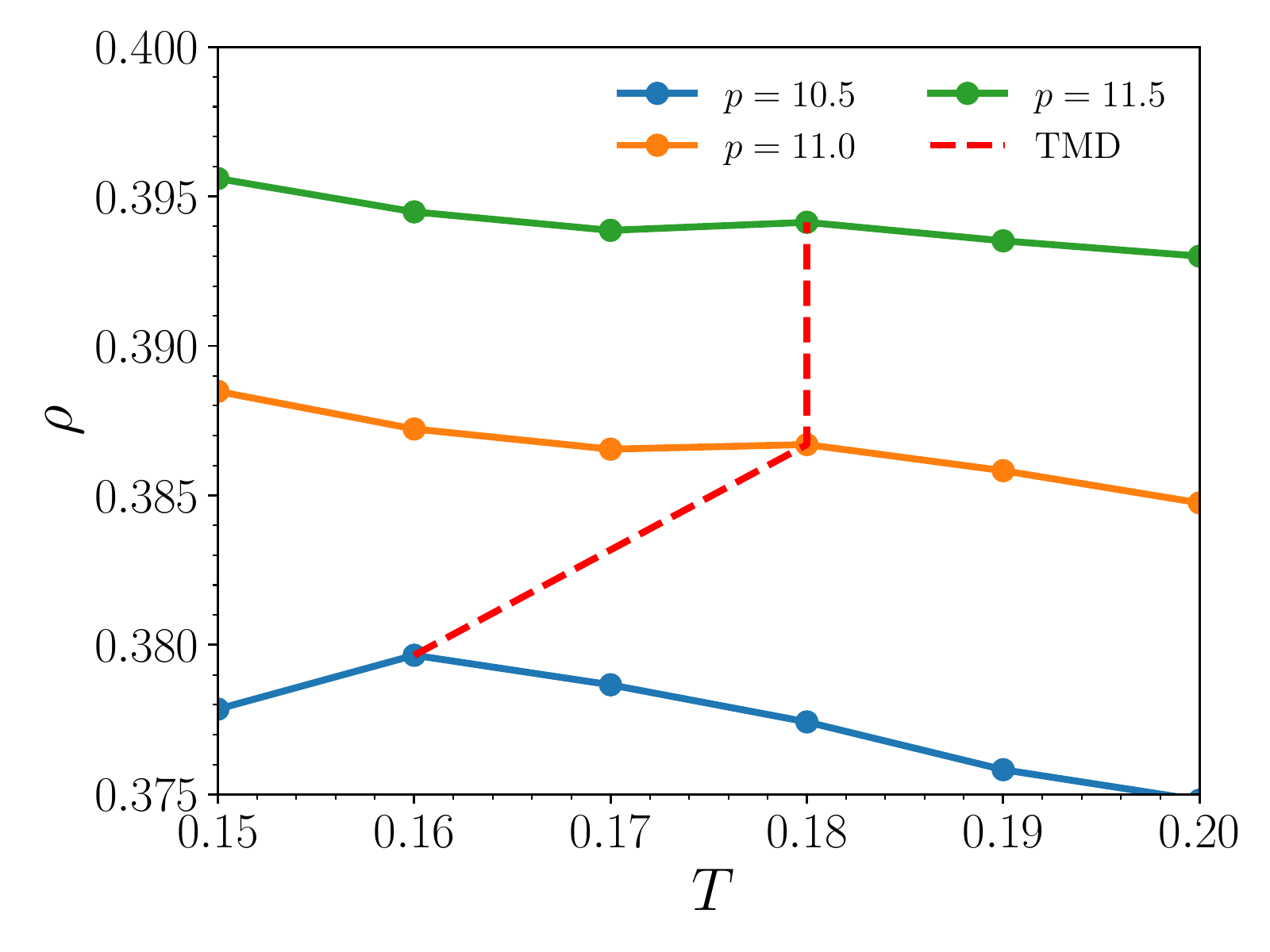}}
    \caption{(a) Diffusion coefficient as function of the system pressure for isotherms where we observed the nematic anisotropic fluid phase. Error bars are smaller than the points on the graph and the lines are for guiding the eyes. (b) The mean-squared displacement (MSD) for the maximum diffusion coefficient observed at $p = 6.5$. Arrow indicates the slope increase as we increase the pressure.}
    \label{fig: diff+tmd}
\end{figure*}

First, we check the dynamic behavior at the fluid phase. The LDT phase melts to the NAL at $p = 5.75$ for the isotherms $0.17 \geq T \leq 0.20$. The dependence of self-diffusion coefficient $D$ with $p$ under compression is shown in Fig.~\ref{fig: diff+tmd} (a). For all isotherms, $D$ has a maxima at $p = 6.50$. It may look curious at a first glance once we expect a decrease in $D$ under compression. The behavior observed here is the so-called diffusion anomaly, observed in water and in many two-length scale model~\cite{Angell2014,stanley2008liquid,Gallo2016,de2008waterlike,nogueira2020tracer}.Also, the density anomaly is characterized by a maxima in the density upon cooling at constant pressure. The Temperature of Maxima Density (TMD) is show in Fig.~\ref{fig: diff+tmd} (b) for the isobars where it was observed. 


\section{Conclusion}\label{sec:conclusion}

In this work we explore the phase diagram of a dumbbell model composed of two HCSC beads with intramolecular separation $\lambda = 0.50$. Such system was chosen since it has show a variety of stripes patterns, with end-to-end, T-like and side-by-side alignments. Our results showed how is the transition between this distinct patterns by analyzing the thermodynamic and structural changes along compression isotherms. Besides the stripes and the LDT solid phases, we have observed a nematic anisotropic phase, with polymer-like pattern, at high temperatures and intermediate pressures. This reentrant fluid phase indicates that the EES and TS patterns are less stable than the SSS pattern, which did not melted for the simulated temperatures. 

Also, we showed how that the new characteristic length scale at $r_1 + \lambda$, which comes from the dumbbell anisotropic geometry, play a major role in the EES-TS-SSS transition. Not only the structural properties have a interesting behavior, but the diffusion in the nematic fluid phase shows an anomalous increase under compression and the density has a temperature of maximum density. Unlike the water anomaly, where the fluid phase has a minima and maxima in $D$, here the minima coincides with the solidification pressure. This results can assist the design nanoparticles based materials with specific mesopatterns.

\section*{Conflicts of interest}
There are no conflicts to declare.

\section*{Acknowledgements}
Without public funding this research would be impossible.
The authors are grateful to the Brazilian National Council for Scientific and Technological Development (CNPq, proc.  407818/2018-9), Coordination for the Improvement of Higher Education Personnel (CAPES, financing Code 001),  Research Support Foundation of the State of Rio Grande do Sul (FAPERGS, TO 21/2551-0002024-5), for the funding support.

\section*{Credit Author Statement}

{\bf{Thiago P. O. Nogueira}}: Methodology, Software, Data Curation Validation, Formal analysis, Investigation, Writing - Original Draft. {\bf{José Rafael Bordin}}: Conceptualization, Methodology, Resources, Writing - Review \& Editing, Supervision, Project administration, Funding acquisition.



 \bibliographystyle{elsarticle-num} 
 \bibliography{ref}





\end{document}